\documentclass[a4paper,12pt]{article}

\usepackage{amsmath}
\usepackage{amssymb}
\usepackage{amsthm}
\usepackage{graphicx}
\usepackage{xcolor}
\usepackage{url}
\usepackage{subcaption}
\usepackage{siunitx}
\usepackage{algorithm}
\usepackage{algpseudocode}
\usepackage{a4wide}
\usepackage[T1]{fontenc} 

\algnewcommand{\LineComment}[1]{\State \(\triangleright\) #1}
\DeclareMathOperator*{\argmin}{arg\,min}
\newcommand{\R}{\mathbb{R}}
\newcommand{\figsize}{0.75\textwidth}

\title{Digital twins for city simulation: Automatic, efficient, and robust mesh
  generation for large-scale city modeling and simulation}
\author{Vasilis Naserentin and Anders Logg}

\begin{document}
\maketitle
\markright{\textsc{Mesh generation for large-scale city modeling and simulation}}

\begin{abstract}
  The concept of creating digital twins, connected digital models of
  physical systems, is gaining increasing attention for modeling and
  simulation of whole cities. The basis for building a digital twin of
  a city is the generation of a 3D city model, often represented as a
  mesh. Creating and updating such models is a tedious process that
  requires manual work and considerable effort, especially in the
  modeling of building geometries. In the current paper, we present a
  novel algorithm and implementation for automatic, efficient, and
  robust mesh generation for large-scale city modeling and
  simulation. The algorithm relies on standard, publicly available
  data, in particular 2D cadastral maps (building footprints) and 3D
  point clouds obtained from aerial scanning. The algorithm generates
  LoD1.2 city models in the form of both triangular surface meshes,
  suitable for visualisation, and high-quality tetrahedral volume
  meshes, suitable for simulation.  Our tests demonstrate good
  performance and scaling and indicate good avenues for further
  optimization based on parallelisation. The long-term goal is a
  generic digital twin of cities volume mesh generator that provides
  (nearly) real-time mesh manipulation in LoD2.x.
\end{abstract}
\section{Introduction}

\subsection{Digital twins and 3D city models}

The need for creating digital replicas of physical systems has been
around for decades~\cite{ketzlerDigitalTwinsCities2020} and the term
\textit{twin} was introduced already by NASA's Apollo
program~\cite{liuReviewDigitalTwin2021}. In tandem, the term Digital Twin (DT) has been
attracting significant attention in the last few years, especially in
the context of the urban
environment~\cite{ketzlerDigitalTwinsCities2020}. Behind the
contemporary term lies mostly what software engineers refer to as
\textit{plain old data}, derived from multiple sources, stitched
together in various forms and ready for consumption for different
needs of the end-users of the twin. On the city scale, since the DT is
required to serve a variety of purposes, including multi-physics
simulations~\cite{ranjbar3DAnalysisInvestigation2012,rivasCFDModellingAir2019}, what-if
scenarios~\cite{bahu3DSpatialUrban2014,galimshinaWhatOptimalRobust2021} and life-cycle
analysis~\cite{strzalka3DCityModeling2011,heiselHighresolutionCombinedBuilding2022}, a common groundwork for the
different means of the consumption of data is of great importance;
even if the digital twin needs to adapt to which data is being
analysed, the common denominator is a 3D city model~\cite{pettitVISUALISATIONSUPPORTEXPLORING}.

3D city models generally enclose buildings and other built objects
(road networks, green areas etc.). Often the process of 3D model
creation is a manual or semi-manual process that is both tedious,
time-consuming, and error-prone. Moreover, man-made city models tend
to become outdated since their creation is a one-time project. Hence,
creating and maintaining DTs is a challenging
task~\cite{yao3DCityDB3DGeodatabase2018,ledouxCityJSONCompactEasytouse2019,ohoriModelingCitiesLandscapes2018,biljeckiMOSTCOMMONGEOMETRIC2016,ledoux3dfierAutomaticReconstruction2021}.
The desired outcome is an automatically generated detailed 3D city
model, or even multiple 3D models serving various analysis needs and
purposes. The input to the process can be aerial imagery, topographic
data, 3D LiDAR data, or a fusion of them. This 3D-fication process is
a multi-step procedure and depending on the available input data and
methods used, the quality of the outcome may vary greatly. For a
complete, end-to-end solution, the pipeline involves developing robust
systems for building detection, rooftop recognition, geometry
modeling, and, eventually, mesh generation.

\subsection{Contribution of this work}

In the present work, we tackle the problem of creating
LoD1.2~\cite{biljeckiFormalisationLevelDetail2014} 3D models of any urban
environment in the form of both triangular surface meshes and
tetrahedral volume meshes. The mesh generation is based on publicly
available datasets from the Swedish Mapping, Cadastral and Land
Registration Authority (https://www.lantmateriet.se) or any other similar repository
that adheres to the same standards.

Our mesh generation strategy is unique in that it generates not only
triangulated surfaces but also \emph{conforming tetrahedral volume
  meshes}, suitable for high-performance simulations of, e.g., urban
wind comfort and air quality by computational fluid dynamics
(CFD). Such simulations require the 3D model generation to not only
provide a surface model but also a high-quality computational volume
mesh that is watertight, conforming, and satisfies a certain minimal
angle condition.

The long-term goal is the creation of an automated process by which
the computational mesh is generated as part of an \emph{interactive}
simulation environment. The demand for interactivity implies that the
mesh generation be both \emph{efficient}, i.e., the mesh generation
should require minimal time as part of a simulation pipeline, and
\emph{robust}, i.e., the mesh generation pipeline should never break
down as a result of bad or unexpected input data. The algorithms and
software presented in the present work partially solve these goals and
avenues for future optimisations are discussed to reach the long-term
goal of (near) real-time generation of 3D models and meshes.

Our implementation DTCC Builder~\cite{loggDTCCBuilder2022} is part of
the Digital Twin Platform developed at the Digital Twin Cities Centre
(\url{https://dtcc.chalmers.se}) and is available as free/open-source
software (under the permissive MIT license).

\subsection{Related work}

The topic of automated creation of 3D city models from raw input data
has received increasing attention in recent years~\cite{wangLiDARPointClouds2018}, and
there are many examples of 3D city models being created for specific
consumption. For example in~\cite{buyukdemircioglu3DCampusApplication2018}, a 3D model of
the Beytepe Campus of Hacettepe University was generated and
visualized using both Cesium and Potree JavaScript libraries. The
authors initially used stereo photographs from an airplane which
helped them generate a dense point cloud. Then, the point cloud was
converted to a Digital Surface Model (DSM). The DSM was complemented
by manually generating 106 building footprints from the stereo images.
Finally, the DSM was combined with the building footprints in order to
generate 3D building geometries in LoD1.2.

In~\cite{tschirschwitzDuisburg1566Transferring2019}, a 3D model of the city of Duisburg in the
year 1566 was generated. To generate the 3D model of the city in that
particular year, the authors used close-range terrestrial laser
scanning on a wooden model of the city that exists in a cultural
historic museum. The point cloud created by the scanning was combined
with ortho-images of the wooden model to create 3D models of the
buildings. Another example using LiDAR data is presented in
\cite{prietoApplicationLiDARData2019} for analysis of the solar potential of building
roofs in an urban 3D model.

In the same spirit, Lindberg \textit{et al.}~\cite{lindbergSolarEnergyBuilding2015}
present a model for estimating shortwave irradiance on ground, roofs
and building walls. To showcase the results, the authors generated a 3D
visualisation of an area in Gothenburg, Sweden. The ground and
building DSMs were derived using geodata; two sets of data were used
to derive the DSM, one LiDAR dataset for the ground heights and one
high level of detail 3D vector layer describing the roof structures.
Additional LiDAR datasets were used to generate a canopy DSM
representing the height of bushes and/or trees and then the authors
used the OpenGL library in R to showcase the results. Jian and
Fan~\cite{jianThreedimensionalVisualizationHarmful2014} present a simulation of harmful gas diffusion in
3D environment of an urban area. The authors performed their
simulations in a generated model provided by Digital Earth Scientific
Platform~\cite{huadongDIGITALEARTHSCIENCE}.

Striving to provide a more generic approach, the authors
of~\cite{rajpriyaGeneration3DModel2014} present a workflow for
generating 3D models for any city. The researchers created a 3D city
model in LoD1 for a part of the city of Ahmedabad in India. Initially,
the contributors created DEMs using photogrammetry from satellite
stereo images. Then, Automatic Terrain Extraction (ATE) was used to
automatically extract DTMs from the generated DEMs. To automatically
extract building footprints from high resolution satellite imagery, the
authors used a combination of commercial software, Arc GIS and
eCognition. In~\cite{nagelerNovelValidatedMethod2017}, a novel
validated method for GIS based automated dynamic urban building energy
simulations is presented. The authors visualize the city of Gleisdorf
in 3D to highlight power consumption of all buildings. GIS data were
collected to visualize the 2D building geometry. Then, a 2.5D geometry
was generated from the 2D geometry and the height of the laser
scanning for each building. Afterwards, building heights were assigned
to each building directly in QGIS.

In~\cite{chenIFC3DTiles}, a different approach is used whereby the 3D city
model is created from a Building Information Model (BIM)
\cite{cerovsekReviewOutlookBuilding2011}. The authors present a workflow going from BIM to
a 3D city model, but rely on the input data being handmade Industry
Foundation Classes (IFCs). The authors conclude that the process is
both time consuming and resource expensive, emphasizing the complexity
of 3D city model creation. Alomia \textit{et al.}~\cite{alomiaAutomatic3DUrban2019}
present the integration of procedural modeling and GIS for the
development of a 3D city model. The workflow consists of geographic
data capture of the area of interest on which building footprints are
identified. Then, a mesh extrusion takes places to generate LoD1
buildings.

In~\cite{prechtelStrategiesAutomationUpgrading2015}, a technique for upgrading large-scale 2D GIS
landscape representations into schematic 3D models is presented. The
workflow uses a readily available 2D vector model along with elevation
data. Geometric objects are generated from geo-processing tools and
footprint shape extrusion results in LoD1 buildings.  Additionally,
the authors enriched the resulting model with some LoD2 assets that
were modeled and photo-textured. In~\cite{richterConceptsTechniquesIntegration2014} a system that
is capable of integrating, analysing and visualising 3D point clouds
is presented. The implementation also provides pre-processing and
analysis features to perform object class segmentation.
In~\cite{tobiasRapidReconstructionHistorical2018}, a workflow to create a landscape model and 3D
buildings based on old maps, plans, drawings and photographs is
presented. The authors combined GIS techniques, 3D CAD software and
procedural modelling tools.

In 2021, Ledoux \textit{et al.} released open-source
software~\cite{ledouxCityJSONCompactEasytouse2019,ledoux3dfierAutomaticReconstruction2021}
that allows the automated 3D reconstruction of city models by using 2D
topographical datasets which elevates them to the height obtained by
point clouds. The end result is, according to the developers, ``\ldots
surface(s) that aim(s) to be error-free: no self-intersections, no
gaps, etc''. The same group very recently extended their work and
released a new piece of open-source software
City4CFD~\cite{padenAutomaticReconstruction3D2022}. The code
reconstructs terrain, buildings and surface layers (including water,
roads, and vegetation) by combining 2D geographical datasets and LiDAR
elevation data. The resulting geometry is then used as input to an
external unstructured finite volume mesh generator and is used for CFD
simulations using the OpenFOAM
package~\cite{wellerTensorialApproachComputational1998}. The end-goal
of these contributions is creating 3D models that can be used for more
advanced analysis and simulation and not only for visualisation. This
goal is shared by the present work.

\subsection{Outline of this paper}

In the remainder of this paper, we first present the results of our
work on an automated process for efficient and robust 3D city model
generation. The results consist of an overview of the algorithm and
software, followed by a benchmark study on both synthetic and
real-world data. We next discuss the performance and limitations of
the algoritm and software based on the results the benchmark study. We
then move on to presenting in some detail the algorithm for 3D city
model generation, consisting of a pipeline of preprocessing and mesh
generation steps. Finally, we comment on the implementation and how to
obtain the software.

\section{Results}

\subsection{Overview of algorithm and software}

The starting point for our mesh generation pipeline is a cadastral
map, i.e., a set of 2D polygonal building footprints, and a
corresponding 3D point cloud, i.e., a set of 3D points. The cadastral
map is assumed to be given in the standard shapefile (\url{.shp})
format and the point cloud is assumed to be given in the standard LAS
(\url{.las} or \url{.laz}) format. Such data are readily available for
most cities and in particular for the whole of Sweden from the Swedish
Mapping, Cadastral and Land Registration Authority where we have
sourced our data.

Given these two datasets, in combination with a 3D bounding box, our
algorithm generates a tetrahedral mesh of the volume defined by the
intersection of the bounding box and the empty space above and between
the buildings and the ground. From the tetrahedral volume mesh, we
also extract a triangular surface mesh by computing the boundary of
the tetrahedral volume mesh. This mesh is automatically watertight and
boundary conforming in the sense that the triangles on building
surfaces necessarily align with the triangle surface (TIN)
representing the ground.

\begin{figure}[htb]
  \centering
  \includegraphics[width=\figsize]{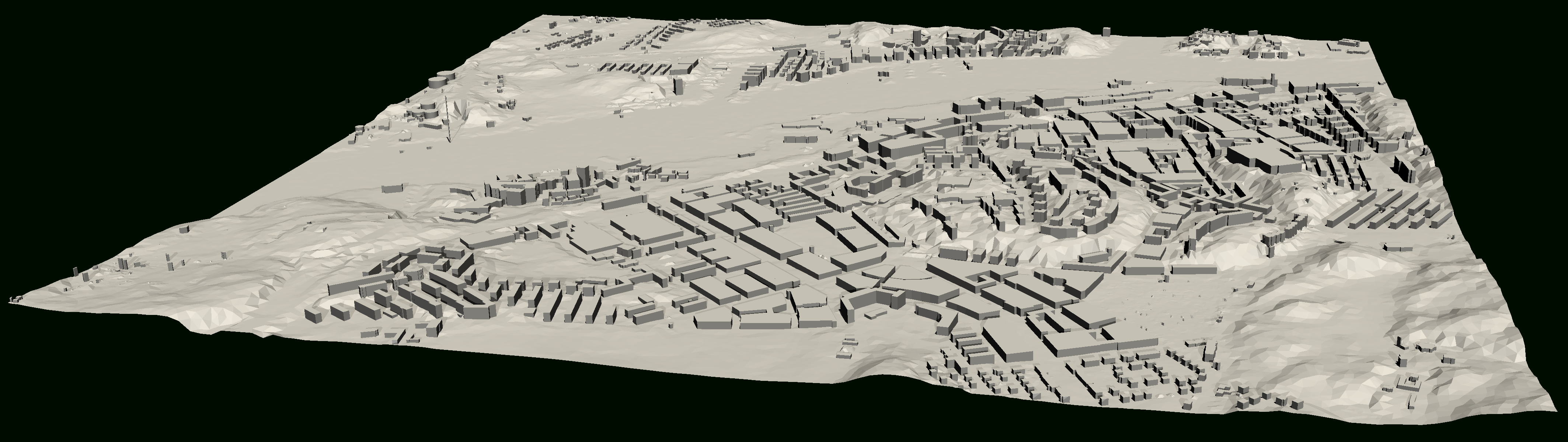} \\
  \includegraphics[width=\figsize]{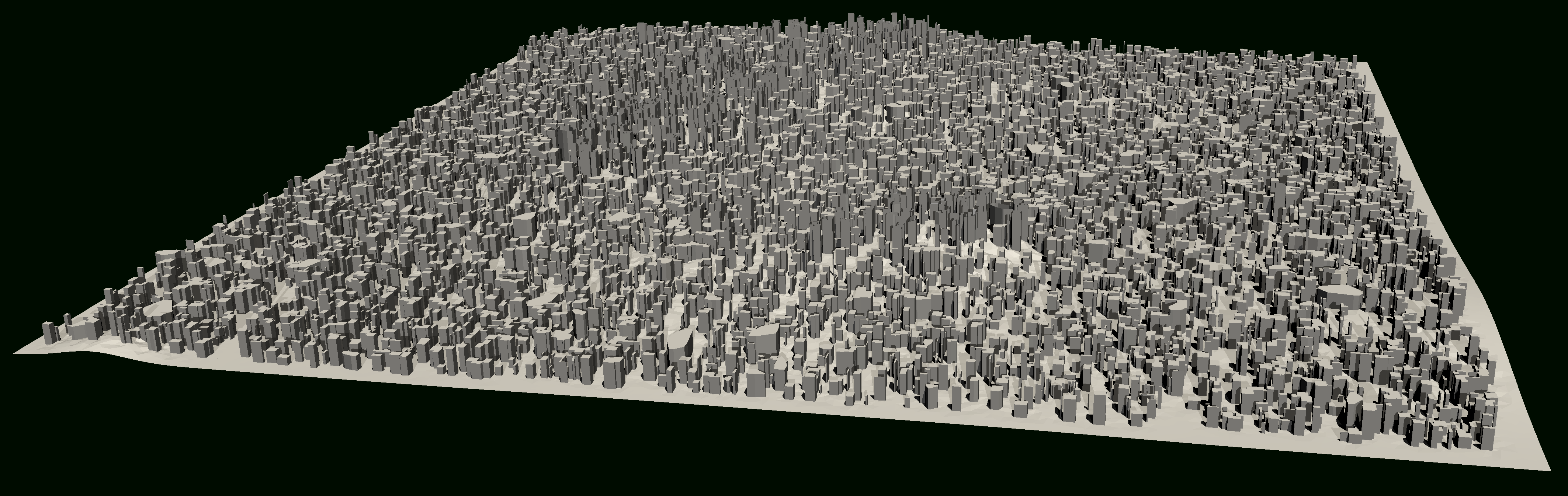}
  \caption{Top: Surface mesh of Majorna in Gothenburg
    Sweden. The city model consists of \num{3096} buildings.
    Bottom: Surface mesh of a synthetic city model with \num{11585}
    randomly placed buildings. Both models cover an area of size
    \SI{2.5}{\kilo\metre}~$\times$~\SI{2.5}{\kilo\metre}.}
  \label{fig:demo}
\end{figure}

In Figure~\ref{fig:demo}, we demonstrate our algorithm for a pair of
demo datasets. The first dataset named Majorna is a
\SI{2.5}{\kilo\metre}~$\times$~\SI{2.5}{\kilo\metre} large area in
central Gothenburg, Sweden. This dataset is publicly available from
the Swedish Mapping, Cadastral and Land Registration Authority. The
dataset is also available as part of the distribution of our
software~\cite{loggDTCCBuilder2022}. The second dataset is a synthetically
generated (randomized) city model which allows the building density to
be controlled via a parameter. The figure shows the end result of the
mesh generation pipeline as triangular surface meshes obtained by
extracting the boundary of the generated tetrahedral volume
meshes. Note that we show the surface meshes rather than the volume
meshes since the surface meshes are more readily visualised.

Our mesh generation algorithm relies on two key ideas. First, the mesh
generation is reduced from a 3D problem to a 2D problem by taking
advantage of the cylindrical geometry of extruded 2D footprints; a 2D
mesh respecting the boundaries of the buildings is generated by a 2D
mesh generator and then layered to form a 3D mesh. Second, the 3D mesh
is adapted to the geometries of building and ground by solving a
partial differential equation (PDE) with the ground and building
heights as boundary conditions (mesh smoothing). Together these two
ideas enable the creation of a both efficient and robust pipeline for
automated large-scale mesh generation from raw data. Our algorithm is
described in detail in the Methods section of our paper.

\subsection{Benchmark study}

To evaluate the efficiency and detect possible bottlenecks of our mesh
generation pipeline, we conduct a benchmark study consisting of two
different experiments. We investigate (a) how the computational cost
scales with the mesh resolution (number of cells generated) and (b)
how the computational cost scales with the number of buildings for a
fixed area. For (a), we use a public dataset bundle from the Swedish
Mapping, Cadastral and Land Registration Authority of an area located
in Central Gothenburg, Sweden, a district called Majorna. For (b), we
create a synthetic city, by placing $N$ randomly positioned buildings
on a square tile the same size as the Majorna test case. Both
configurations span the same area of \SI{6.25}{\kilo\metre^2}.

All benchmark cases were run on the same compute node powered by a
Xeon\textsuperscript{\tiny\textregistered} Silver 4110 CPU @ 2.10GHz
using a single thread. The maximum memory footprint was approximately
64~GB and 2~GB for the Majorna and synthetic city datasets,
respectively.

\begin{figure}[htb]
  \centering
  \includegraphics[trim={0.75cm 0.75cm 0.75cm 0.75cm},clip,width=\figsize]{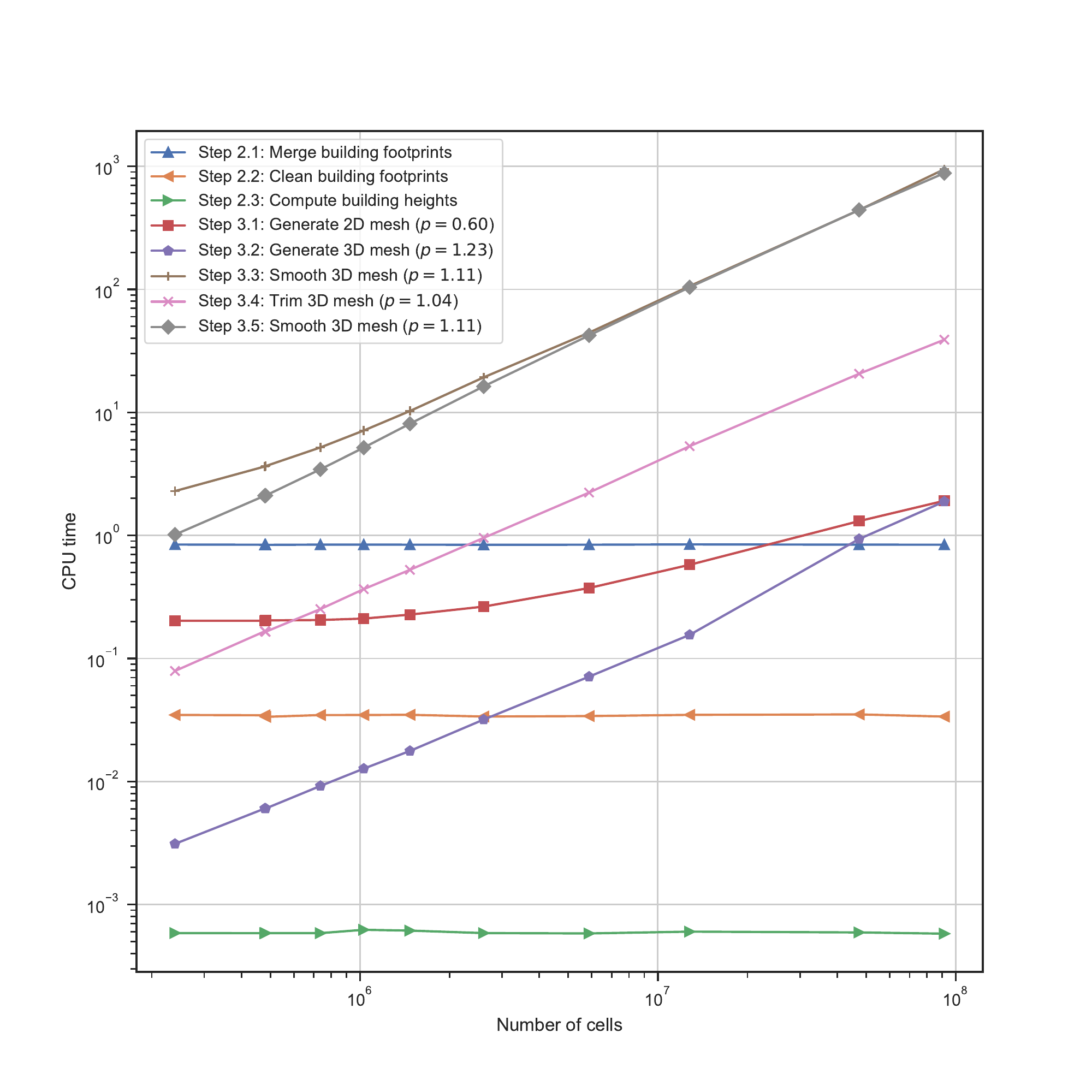}
  \caption{Benchmark study (a): Computational cost (CPU time) as
    function of the number of cells (tetrahedra) generated for the
    different steps of our mesh generation pipeline for the Majorna
    dataset.}
  \label{fig:bench:majorna}
\end{figure}

\begin{figure}[htb]
  \centering
  \includegraphics[trim={0.75cm 0.75cm 0.75cm 0.75cm},clip,width=\figsize]{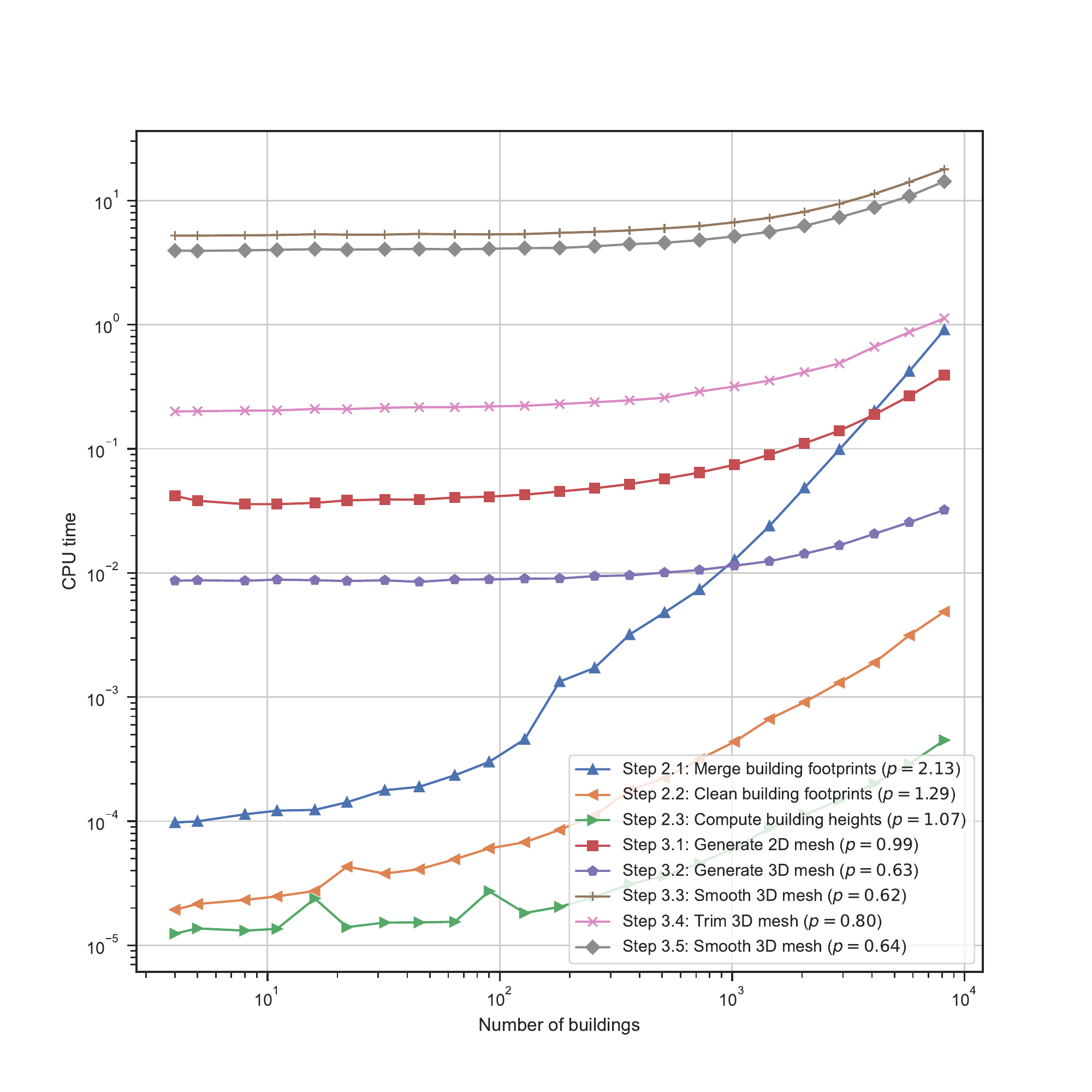}
  \caption{Benchmark study (b): Computational cost (CPU time) as
    function of the number of buildings for the
    different steps of our mesh generation pipelline for the synthetic
    dataset.}
  \label{fig:bench:random}
\end{figure}

The results of the benchmarks study are summarised in
Figure~\ref{fig:bench:majorna} and Figure~\ref{fig:bench:random}. The
different steps (2.1--3.5) refer to the steps in our mesh generation
pipeline described in the Methods section. We only measure the cost of
steps 2.x and 3.x since steps 1.x consist of initial preprocessing.

\section{Discussion}

\subsection{Findings}

Our algorithm and implementation has proven to be generic and robust
for a range of mesh resolutions, building densities, and also data
quality. A substantial amount of work has gone into cleaning and
filtering of data to ensure that the algorithm runs smoothly even for
bad quality data. However, bad input data will lead to unexpected
results. This can be seen in the top left corner of
Figure~\ref{fig:demo} which shows a spike in the generated mesh.  This
spike is the result of a property being located at the site of one of
the \SI{107}{\metre} tall pylons of a bridge (Älvsborgsbron in
Gothenburg, Sweden). This and other cases may be handled by more
extensive filtering of the input data, such as ignoring detected
``buildings'' with aspect ratios larger than some limit.

Looking at the results of the benchmark study in
Figure~\ref{fig:bench:majorna}, we note that, as expected, the
computational cost of Steps~2.1--2.3 is independent of the mesh
resolution since those steps relate the cleaning and extraction of
building heights from the raw data. For Steps~3.2--3.5, we see that
the cost $T = T(N)$ as function of the number of cells is linear in
$N$ ($p = 1$ for $T \sim N^p$). For Step~3.1, the 2D mesh generation
step, the cost scales like $N^{0.6}$, which agrees with linear scaling
in the number of 2D cells generated ($N^{2/3}$). We conclude that the
scaling of all steps of the mesh generation pipeline are optimal.

Turning now to Figure~\ref{fig:bench:random}, the cost as function of
the number of buildings, we see again good scaling for all steps; the
cost of mesh generation is essentially constant as function of the
number of buildings, as long as the city model contains a moderate
number of buildings. When the number of buildings grow and the city
model becomes very dense, the costs start to increase for all
steps. For Steps~2.1--2.3, we expect a linear growth for all
steps. This is also the case, except for Step~2.1 which exhibits an
initial linear growth (when the number of buildings is small to
moderate) and then a quadratic growth (for very dense city
models). The quadratic growth at the extreme end is expected, even if
binning is used (see Algorithm~\ref{alg:MergeCityModel} below) since
at the extreme, each building needs to be merged with all other
buildings. This increase in cost is unproblematic since it only
appears in extreme cases and the cost of merging is dominated by other
costs (see Figure~\ref{fig:bench:majorna}). For Steps~3.1--3.5, the
costs start to grow when the number of buildings becomes large as a
result of the city model becoming more complex and resulting in more
cells being generated.

In summary, the results show optimal scaling in terms of both mesh
resolution and the number of buildings. At the same time, there are
large opportunities for optimisation through parallelism which we
intend to investigate in future work.

\subsection{Limitations}

Substantial work has been put into or mesh generation pipeline.
Nonetheless, there are known limitations that may be addressed in
future work. We comment on some of these limitations here.

\paragraph{Only LoD1.2  supported}

Our mesh generation pipeline currently only supports LoD1.2 city
models (flat roofs). Work on extending the mesh generation pipeline to
LoD1.3 and LoD2.x is in progress, based on a combination of machine
learning for rooftop segmentation from orthophotos and 3D model
generation from pointcloud data. Further extension to LoD3 would
require street level imagery in order to get information about the
building façades~\cite{kolibarovLOD2DigitalTwin2022}.

\paragraph{Lack of parallelism}

Our implementation does not currently support any parallelism. One of
the immediate planned steps is to extensively profile, optimize, and
decide suitable parallelisation strategies for the different steps of
the mesh generation pipeline. Parallelism will open up new
possibilities, e.g., meshing of the whole of Sweden in tiles and
stitching them together, in order to create a watertight mesh on a
national level, similar to the 3D version of the Register of Buildings
and Addresses of the Netherlands (https://3dbag.nl).

\paragraph{Lack of underground meshing}

Our implementation only considers ground and buildings and does
not take into account infrastructure like bridges or green areas. In
addition, we currently only consider mesh generation of the volume
above ground. In future work, we plan to also consider mesh generation
of volumes below ground which is of importance for geotechnical
simulation~\cite{tornborgModellingConstructionLongterm2021}.

\section{Methods}

\subsection{Overview of the mesh generation pipeline}

The meshing pipeline consists of three main steps: (1) city model
generation based on cadastral (2D) data and point cloud (3D) data; (2)
simplification of the city model generated in the first step; and (3)
mesh generation based on the simplified city model from the second
step.

\subsection{Step~1: Generate city model}

In the first step of the mesh generation pipeline, 2D building
footprints are extracted from the cadastral data and building heights
are computed by sampling the point cloud for points that fall inside
the building footprints. This results in a LoD1.2 city model
represented as polygonal prisms $P \times [h_0, h_1]$, where $P$ is
the 2D polygon representing the building, $h_0$ is the ground height
at the centre of the building, and $h_1$ is the absolute computed
height of the building.

The processing pipeline involves several essential steps as outlined
in Table~\ref{tab:step1}. We comment briefly on some of these steps;
full details are provided in the open-source implementation of the
mesh generation pipeline~\cite{loggDTCCBuilder2022}.

\begin{table}[htb]
  \centering
  \small
  \begin{tabular}{|l|l|}
    \hline
    Step~1.1 & Read point cloud data (several LAS/LAZ files) \\
    \hline
    Step~1.2 & Read building footprints (SHP file) \\
    \hline
    Step~1.3 & Compute digital terrain map (DTM) \\
    \hline
    Step~1.4 & Clean building footprints \\
    \hline
    Step~1.5 & Extract building points (from point cloud data) \\
    \hline
    Step~1.6 & Compute building heights (from building points and DTM) \\
    \hline
    Step~1.7 & Export city model \\
    \hline
  \end{tabular}
  \caption{Essential substeps for Step~1: Generate city model.}
  \label{tab:step1}
\end{table}

\paragraph{Step~1.3: Compute digital terrain map (DTM)}

The digital terrain map (DTM) is computed from the point cloud data by
first constructing a regular 2D grid of the domain. The grid size is
configurable and is set by default to \SI{1}{m}. We iterate over all
the points in the point cloud and assign each point to a bin
associated with its closest grid point. We then compute the mean of
the heights ($z$-components) for each bin to set the mean of each grid
point. This step is computationally inexpensive (of linear complexity)
as it only involves iterating once over the point cloud. If the point
cloud is of low quality (sparse), it may happen that the bins are
empty for some grid points. To fill in the values for the missing grid
points, the generation of the DTM is completed by a flood fill
starting from the grid points on the boundary of empty regions. The
generated DTM may be sampled at arbitrary points within the
computational domain and returns the value at the given point by
bilinear interpolation using the four closest grid points (which may
be computed without searching).

\paragraph{Step~1.4: Clean building footprints}

Each building footprint (polygon) is processed to ensure that the
polygons are closed, oriented (counter clock-wise) and that there are
no duplicate vertices or consecutive parallel edges. This step is
necessary since the raw data typically contains a mix of closed and
non-closed polygons with multiple duplicate vertices and edges.

\paragraph{Step~1.5: Extract building points (from point cloud data)}

For each building, we extract all points of the point cloud that fall
within a margin of the the building footprint. This step is
potentially costly since a na\"ive implementation would involve
searching through the entire point cloud for each building. To avoid a
costly search, we build for both the set of building footprints and
the point cloud an axis-algined bounding box (AABB) tree and then
compute the collision between the two AABB trees following standard
algorithms from computational
geometry~\cite{ericsonRealtimeCollisionDetection2004}. The result of
the collision between the two trees is a list of collision candidates;
that is, a list of pairs consisting of a point and a bounding
box. Each such pair signifies a point from the point cloud found to be
located within the bounding box of a building footprint.

Once the collision candidates have been computed, the candidates are
filtered to check whether the points actually fall within the
corresponding building footprint (not just within its bounding box).
This filtering relies on a point-in-polygon algorithm, which is
efficiently implemented based on computing the quadrant
angle~\cite{hormannPointPolygonProblem2001} for each such point in relation to the
polygon.

The filtered points are classified as either ``roof points'' or
``ground points''. Roof points are those that fall within the building
footprint and have LiDAR point classification~6 (Building). Ground
points are those that fall outside the building footprint but closer
than a configurable margin (by default~\SI{1}{m}) to the building
footprint and have LiDAR point classification~2 or~9 (Ground or
Water). Both roof points and ground points are filtered for outliers,
removing all points that fall outside two standard deviations.

\paragraph{Step~1.6: Compute building heights (from building points and DTM)}

Building heights are computed by examining for each building its
ground points and building points from the previous step. The ground
height $h_0$ of the building is set to the 10th percentile of the
ground points and the roof height $h_1$ is set to the 90th percentile
of the roof points. If ground points are missing, then the ground
height $h_0$ is computed by sampling the DTM at the center of the
building footprint. Finally, the generated city model is exported to a
simple JSON format.

\subsection{Step~2: Simplify city model}

The result of Step~1 is a clean LoD1.2 city model consisting of a set of
buildings, each defined by a 2D polygonal footprint, a ground height
and a roof height, together with a digital terrain map that may be
sampled to give the ground height at any point within the
computational domain. However, before a mesh can be generated from the
city model, the city model must be simplified. Simplification is
necessary since building footprints may be located at arbitrarily
small distances or even overlap, which may cause the mesh generation
to either break down or become prohibitively costly by needing to
resolve tiny gaps between buildings. The key steps in the
simplification are outlined in Table~\ref{tab:step2}.

\begin{table}[htb]
  \centering
  \small
  \begin{tabular}{|l|l|}
    \hline
    Step~2.1 & Merge building footprints \\
    \hline
    Step~2.2 & Clean building footprints \\
    \hline
    Step~2.3 & Compute building heights \\
    \hline
    Step~2.4 & Export city model \\
    \hline
  \end{tabular}
  \caption{Essential substeps for Step~2: Simplify city model.}
  \label{tab:step2}
\end{table}

\paragraph{Step~2.1: Merge building footprints}

To prevent very small or even zero distances between buildings, all
pairs of buildings that are closer than a configurable threshold
distance (by default~\SI{1}{m}) are merged into a single building.

The merging process is outlined in
Algorithm~\ref{alg:MergeCityModel}. The algorithm starts from the city
model generated in Step~1 and adds all buildings to a queue. It then
proceeds to pop the first building from the queue and iterates over
all other buildings. If the distance between any such pair of
buildings is found to be closer than the threshold distance
$\varepsilon > 0$, the two building footprints are merged using
Algorithm~\ref{alg:MergePolygons} (described below) and the new
building is pushed to (added to the back of) the queue. The algorithm
proceeds until the queue is empty.

\begin{algorithm}[htb]
  \caption{MergeCityModel (for details, refer to~\cite{loggDTCCBuilder2022})}
  \label{alg:MergeCityModel}
  \small
  \begin{algorithmic}[1]
    \Require{Array $[P_i]_{i=1}^n$ of building footprint polygons}
    \Require{Threshold distance $\varepsilon > 0$}
    \State{Initialize regular grid with resolution $h > \varepsilon$}
    \For{$i \in [1,n]$}
      \State{$\mathcal{B}_i \gets$ bins (grid points) closer than $h$ to $P_i$}
    \EndFor
    \State{Initialize empty $queue$}
    \For{$i \in [1,n]$}
      \State{Push $i$ to $queue$}
    \EndFor
    \While{$queue$ not empty}
      \State{Pop $i$ from $queue$}
      \State{$\mathcal{J} \gets \{j \in [1,n] \;|\; \mathcal{B}_i\cap\mathcal{B}_j \neq \emptyset\}$}
      \For{$j \in \mathcal{J}$}
        \If{$i \neq j$ and $P_j$ not empty}
          \If{$\|P_i - P_j\| < \varepsilon$}
            \State{$P_i \gets $ MergePolygons($P_i$, $P_j$)}
            \State{Push $i$ to $queue$}
            \State{$\mathcal{B}_i \gets$ bins (grid points) closer than $h$ to $P_i$}
            \State{$P_j \gets \emptyset$}
          \EndIf
        \EndIf
      \EndFor
    \EndWhile
  \end{algorithmic}
\end{algorithm}

A simple binning strategy is used to avoid the otherwise quadratic
cost associated with searching all buildings for
overlap/closeness. The binning is based on covering the computational
domain by a regular grid and associating each building (polygon
footprint) to all bins (grid points) such that the (bounding box of)
the polygon is closer than the grid size $h$. This ensures that two
buildings closer than the threshold $\varepsilon > 0$ must always be
associated with a common bin (grid point), as long as
$h > \varepsilon$. Particular care is required in the binning
implementation to ensure that the binning works correctly in the face
of numerical rounding errors. As long as $h > \varepsilon$, the grid
size is arbitrary and numerical experiments indicate that a factor
$1$~to~$4$ times the mean building footprint diameter is a good
choice. Our current implementation uses the factor~$4$.

Algorithm~\ref{alg:MergePolygons} takes two building footprints
(polygons) and returns a single, simple polygon by heuristically
merging the two polygons. Note that only in the case of two
intersecting polygonal domains is it clear what shape the merging
algorithm should create, namely the (polygonal boundary of the) union
of the two polygonal domains. In all other cases, if the polygons do
not intersect, a heuristic choice must be made as to how a new shape
should be created from the two building
footprints. Algorithm~\ref{alg:MergePolygons} first attempts to
compute the union of the two polygons.\footnote{Implemented using the
  GEOS function \texttt{GEOSUnionPrec()}.} If the resulting shape is
acceptable, the union is returned; a shape is acceptable if it is
a simple polygon and the following two conditions are met: (i) no two
distinct vertices are separated by less than the
threshold~$\varepsilon$ and (ii) no vertex is closer
than~$\varepsilon$ to an edge of which it is not an
endpoint.\footnote{Implemented using the GEOS function
  \texttt{GEOSMinimumClearance()}.} If the union is not acceptible,
the algorithm projects the vertices of each of the two polygons onto
the edges of the other polygon (only for vertices that are deemed
close to the other polygon); see
Algorithm~\ref{alg:VertexProjections}. The convex hull of those
projected vertices defines a patch $P$ and the union $A \cup B \cup P$
is then tested for acceptance in the same way. If the patch is not
accepted, then the patch is grown iteratively (by including more
vertices). If this process fails, the algorithm checks the union of
the convex hulls of both polygons, and in the worst-case scenaro, the
convex hull of the two polygons is used.

Algorithms~\ref{alg:MergeCityModel}--\ref{alg:VertexProjections} are
implemented as part of DTCC Builder~\cite{loggDTCCBuilder2022}. For
geometric predicates and set operations, the GEOS is a C/C++ library
is used.~\cite{geoscontributorsGEOSCoordinateTransformation2021}

\begin{algorithm}[htb]
  \caption{MergePolygons (for details, refer to~\cite{loggDTCCBuilder2022})}
  \label{alg:MergePolygons}
  \small
  \begin{algorithmic}[1]
    \Require{Polygon $A$ and polygon $B$}
    \Require{Threshold distance $\varepsilon > 0$} \Comment{Default value used is $\varepsilon = 1\mathrm{m}$}
    \Require{Maximum iterations $N > 0$} \Comment{Default value used is $N = 3$}
    \State{$C \gets A \cup B$} \Comment{Compute union $C$ of $A$ and $B$}
    \If{IsValid($C$, $\varepsilon$)} \Comment{Check if $C$ should be accepted}
      \State\Return{$C$}
    \EndIf
    \For{$k \in [1,N]$} \Comment{Iterate with increasing threshold}
      \State{$P_A \gets \text{VertexProjections($A$, $B$, $\varepsilon$)}$} \Comment{Compute projection of vertices of $A$ onto $B$}
      \State{$P_B \gets \text{VertexProjections($B$, $A$, $\varepsilon$)}$} \Comment{Compute projection of vertices of $B$ onto $A$}
      \State{$P \gets$ ConvexHull($P_A \cup P_B$)} \Comment{Compute convex hull $P$ of vertex projections}
      \State{$C \gets A \cup B \cup P$} \Comment{Compute union $C$ of $A$, $B$ and $P$}
      \If{IsValid($C$, $\varepsilon$)} \Comment{Check if $C$ should be accepted}
        \State\Return{$C$}
      \EndIf
      \State{$\varepsilon \gets 2\varepsilon$} \Comment{Increase tolerance}
    \EndFor
    \State{$C \gets \text{ConvexHull($A$)} \cup \text{ConvexHull($B$)}$} \Comment{Compute union $C$ of convex hulls}
    \If{IsValid($C$, $\varepsilon$)} \Comment{Check if $C$ should be accepted}
      \State\Return{$C$}
    \EndIf
    \State{$C \gets \text{ConvexHull($A \cup B$)}$}
    \Comment{Fall back to convex hull of union}
    \State\Return{$C$}
  \end{algorithmic}
\end{algorithm}

\begin{algorithm}[htb]
  \caption{VertexProjections (for details, refer to~\cite{loggDTCCBuilder2022})}
  \label{alg:VertexProjections}
  \small
  \begin{algorithmic}[1]
    \Require{Polygon $A$ and polygon $B$}
    \Require{Threshold distance $\varepsilon > 0$} \Comment{Default value used is $\varepsilon = 1\mathrm{m}$}
    \State{$P \gets \emptyset$} \Comment{Initialize empty set of points}
    \For{$v \in \mathrm{Vertices}(A)$} \Comment{Iterate over vertices of $A$}
    \For{$e \in \mathrm{Edges}(B)$} \Comment{Iterate over edges of $B$}
    \State{$p \gets \argmin_{q\in e} \|v - q\|$} \Comment{Compute closest point on edge}
    \If{$\|v - p\| < \varepsilon$} \Comment{Check if point is close enough}
    \State{$P \gets P \cup \{p\}$} \Comment{Add point}
    \EndIf
    \EndFor
    \EndFor
  \end{algorithmic}
\end{algorithm}

\begin{figure}[htb]
  \centering
  \begin{subfigure}[t]{0.475\textwidth}
    \centering
    \includegraphics[width=\figsize]{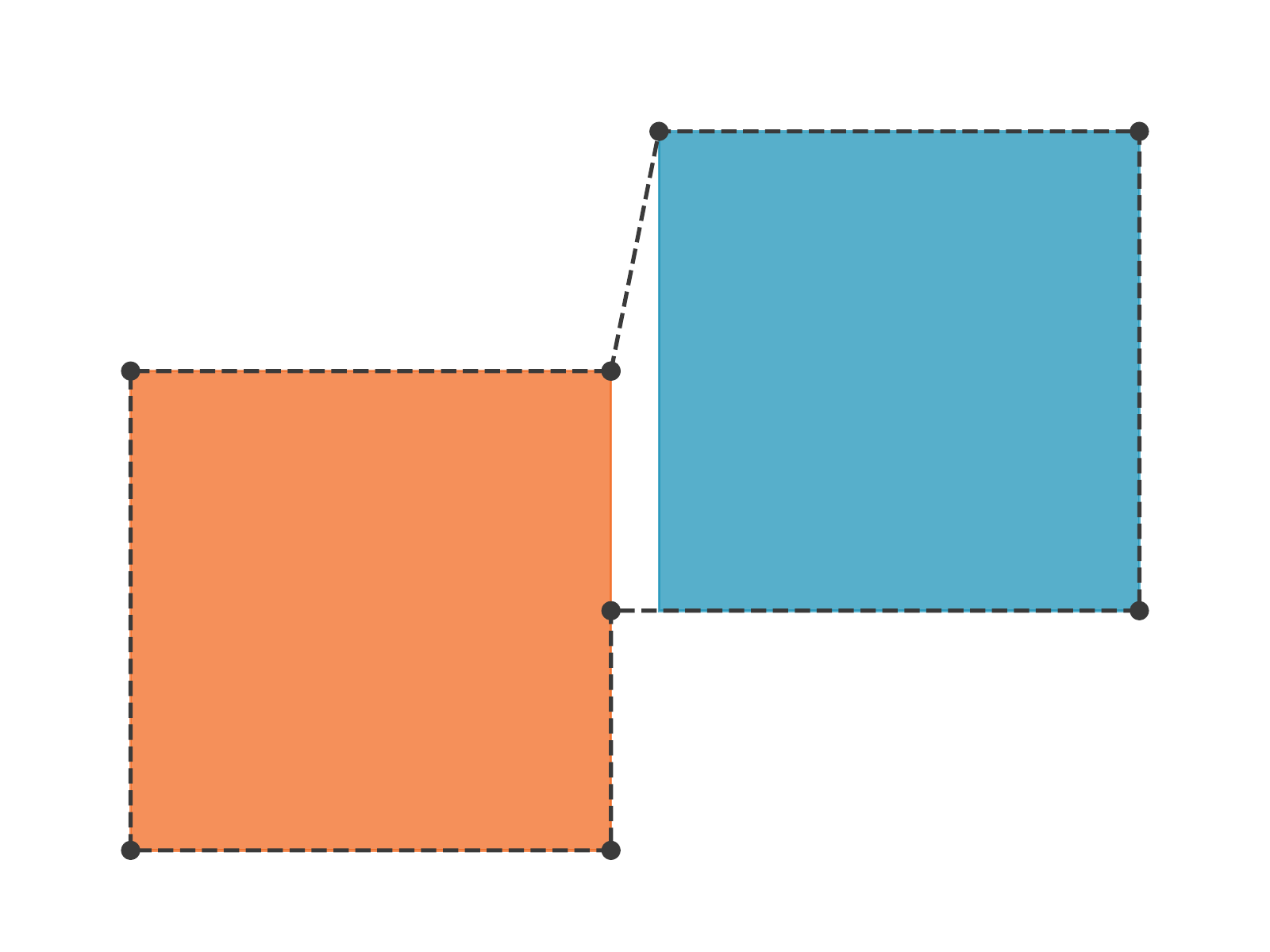}
    \label{fig:MergePolygons:simple:0}
  \end{subfigure}
  \begin{subfigure}[t]{0.475\textwidth}
    \centering
    \includegraphics[width=\figsize]{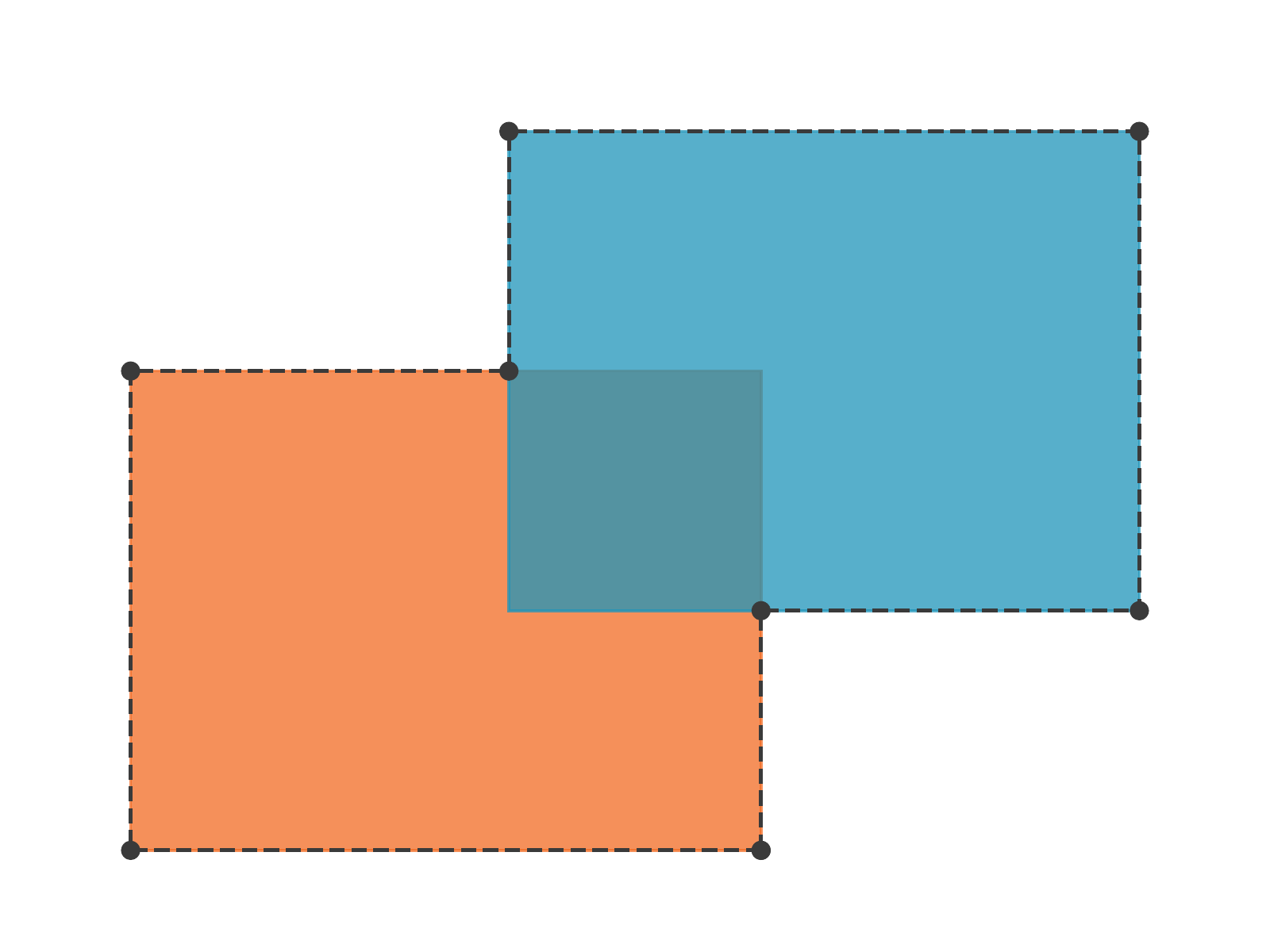}
    \label{fig:MergePolygons:simple:1}
  \end{subfigure}
  \begin{subfigure}[t]{0.475\textwidth}
    \centering
    \includegraphics[width=\figsize]{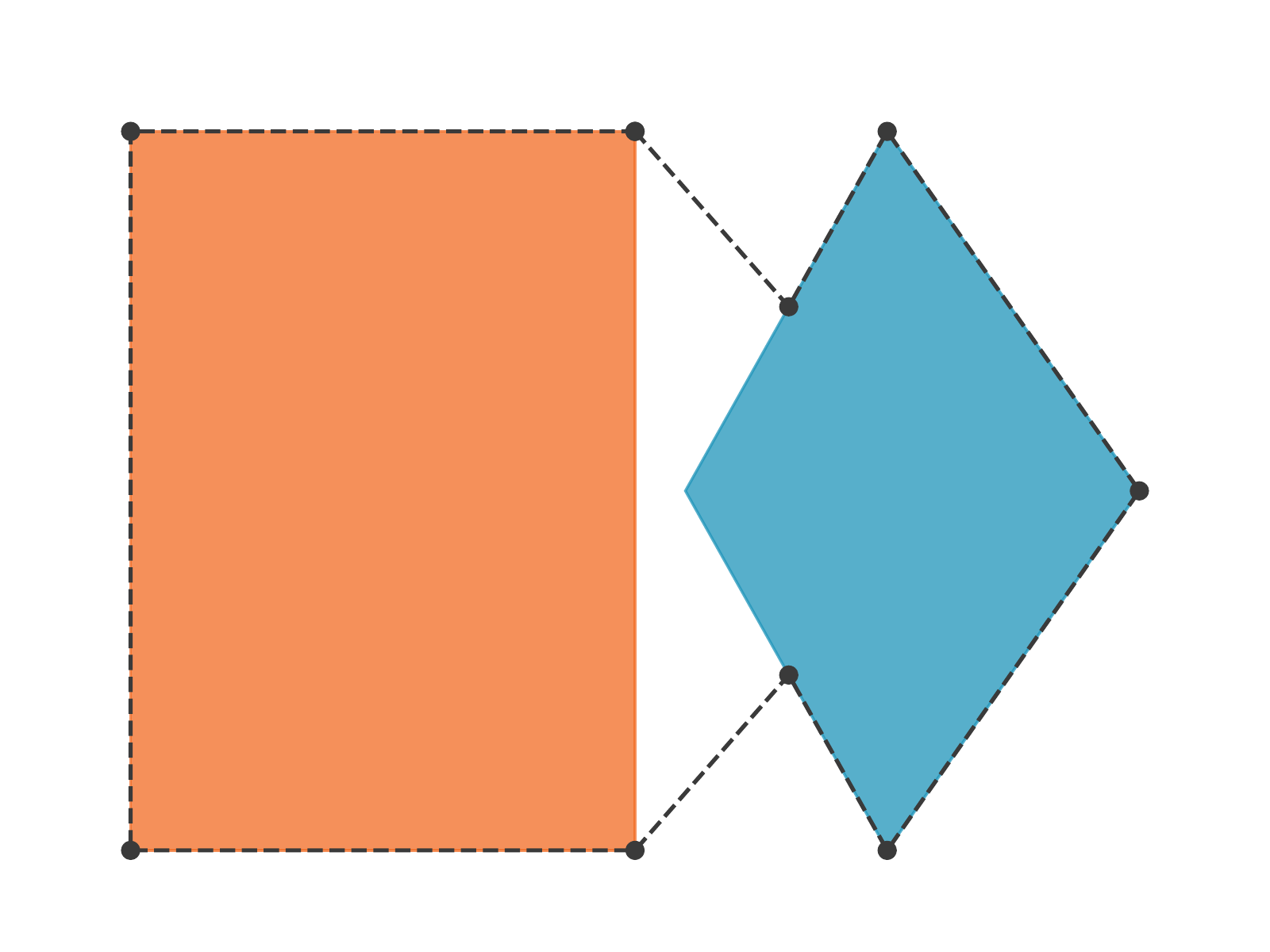}
    \label{fig:MergePolygons:simple:2}
  \end{subfigure}
  \begin{subfigure}[t]{0.475\textwidth}
    \centering
    \includegraphics[width=\figsize]{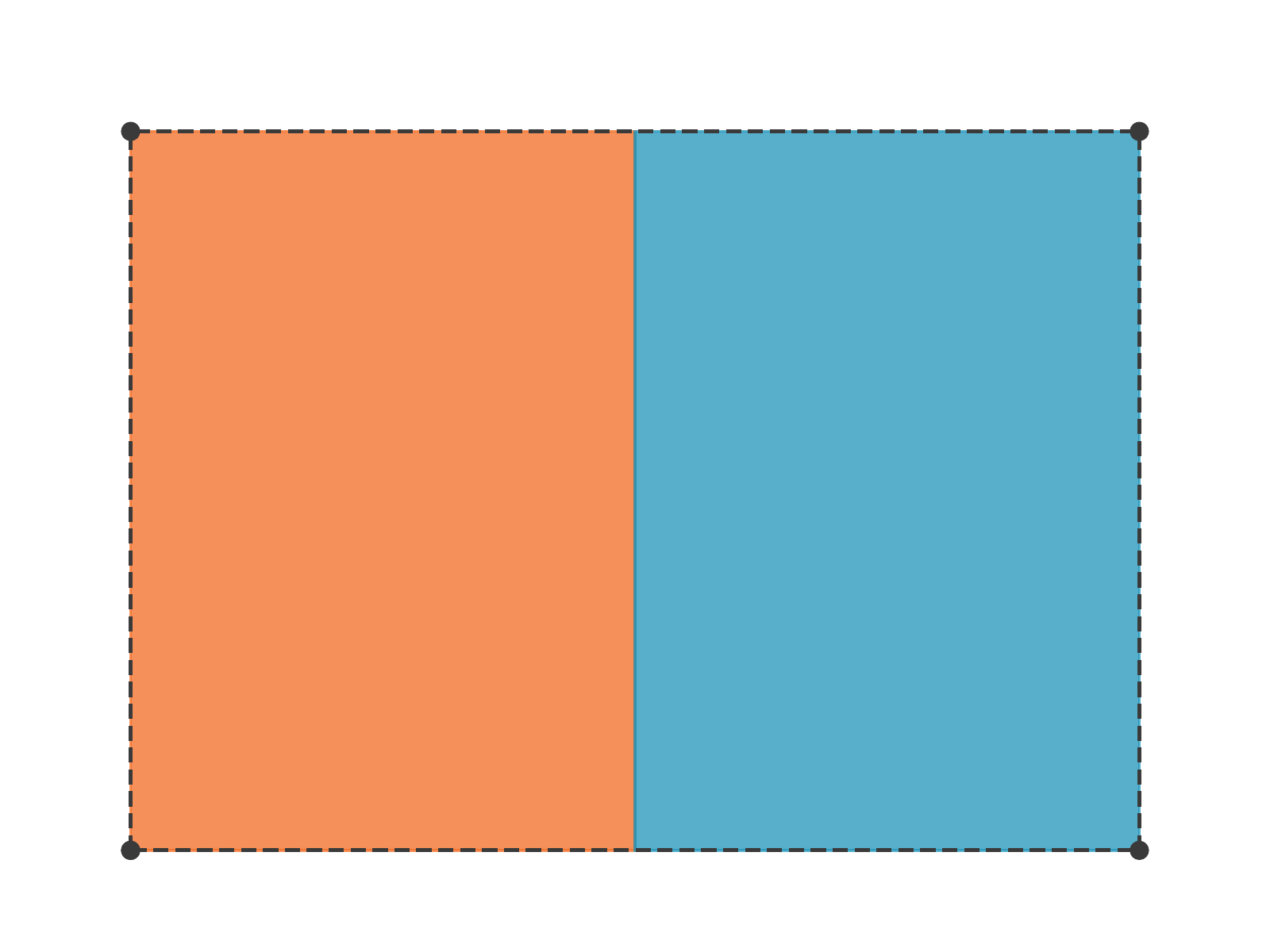}
    \label{fig:MergePolygons:simple:3}
  \end{subfigure}
  \caption{Examples of merged building footprints generated by
    Algorithm~\ref{alg:MergePolygons} for a selection of simple test
    cases. In each panel, the two colored shapes show the two polygons
    to be merged and the dashed line and marked vertices show the
    result of the merging algorithm.}
  \label{fig:MergePolygons:simple}
\end{figure}

\begin{figure}[htb]
  \centering
  \begin{subfigure}[t]{0.475\textwidth}
    \centering
    \includegraphics[width=\figsize]{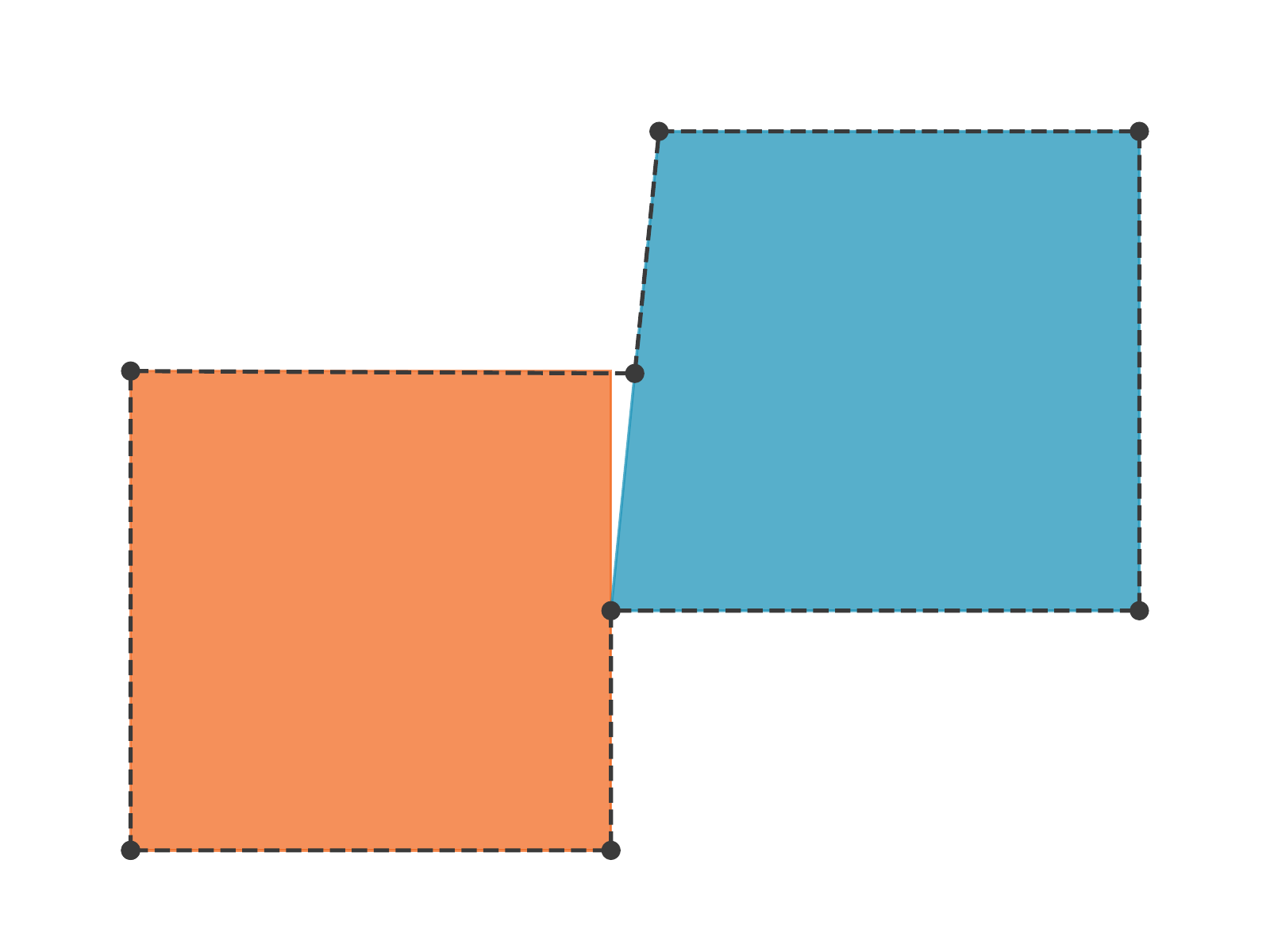}
    \label{fig:MergePolygons:simple:5}
  \end{subfigure}
  \begin{subfigure}[t]{0.475\textwidth}
    \centering
    \includegraphics[width=\figsize]{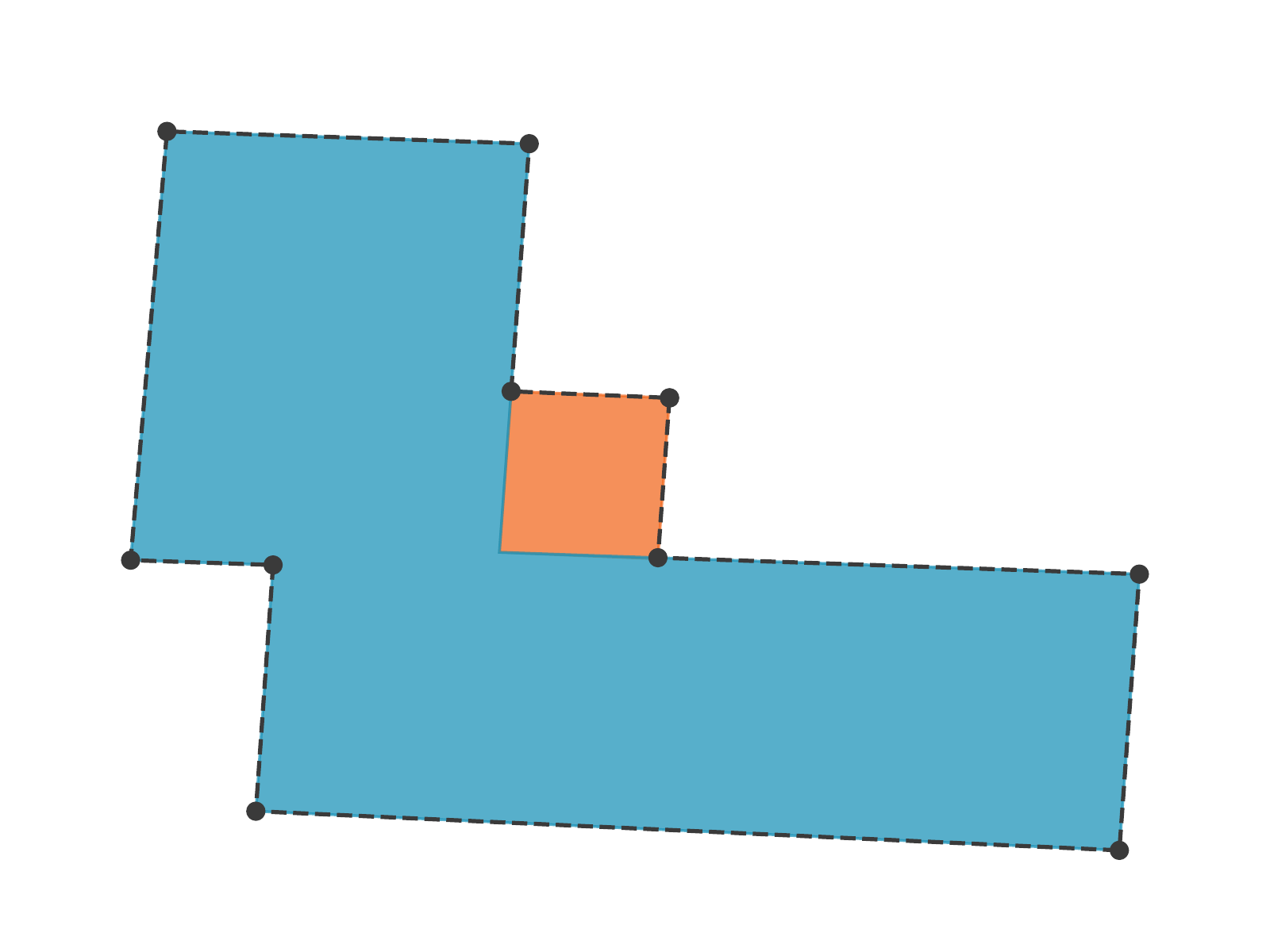}
    \label{fig:MergePolygons:simple:9}
  \end{subfigure}
  \begin{subfigure}[t]{0.475\textwidth}
    \centering
    \includegraphics[width=\figsize]{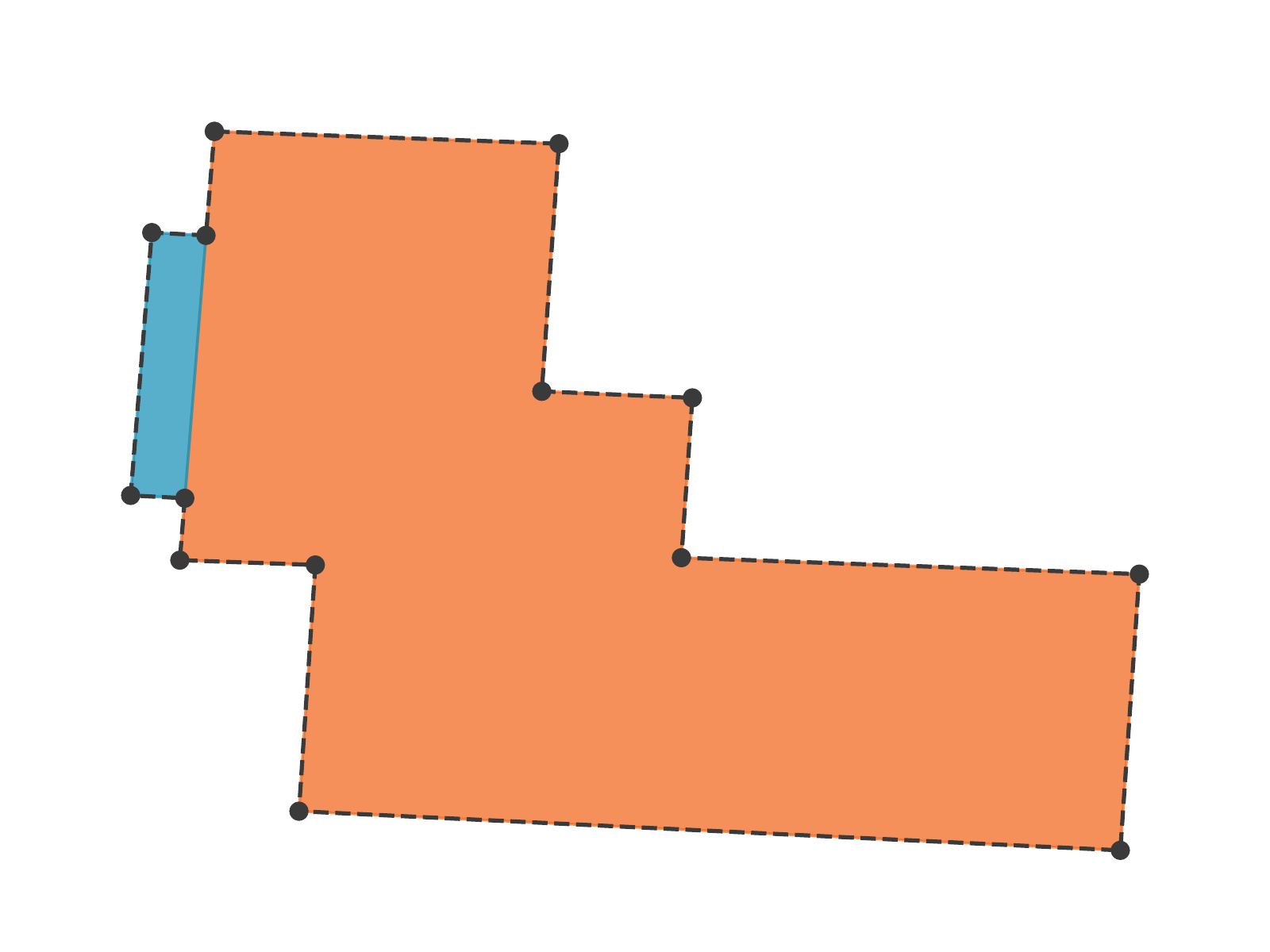}
    \label{fig:MergePolygons:simple:10}
  \end{subfigure}
  \begin{subfigure}[t]{0.475\textwidth}
    \centering
    \includegraphics[width=\figsize]{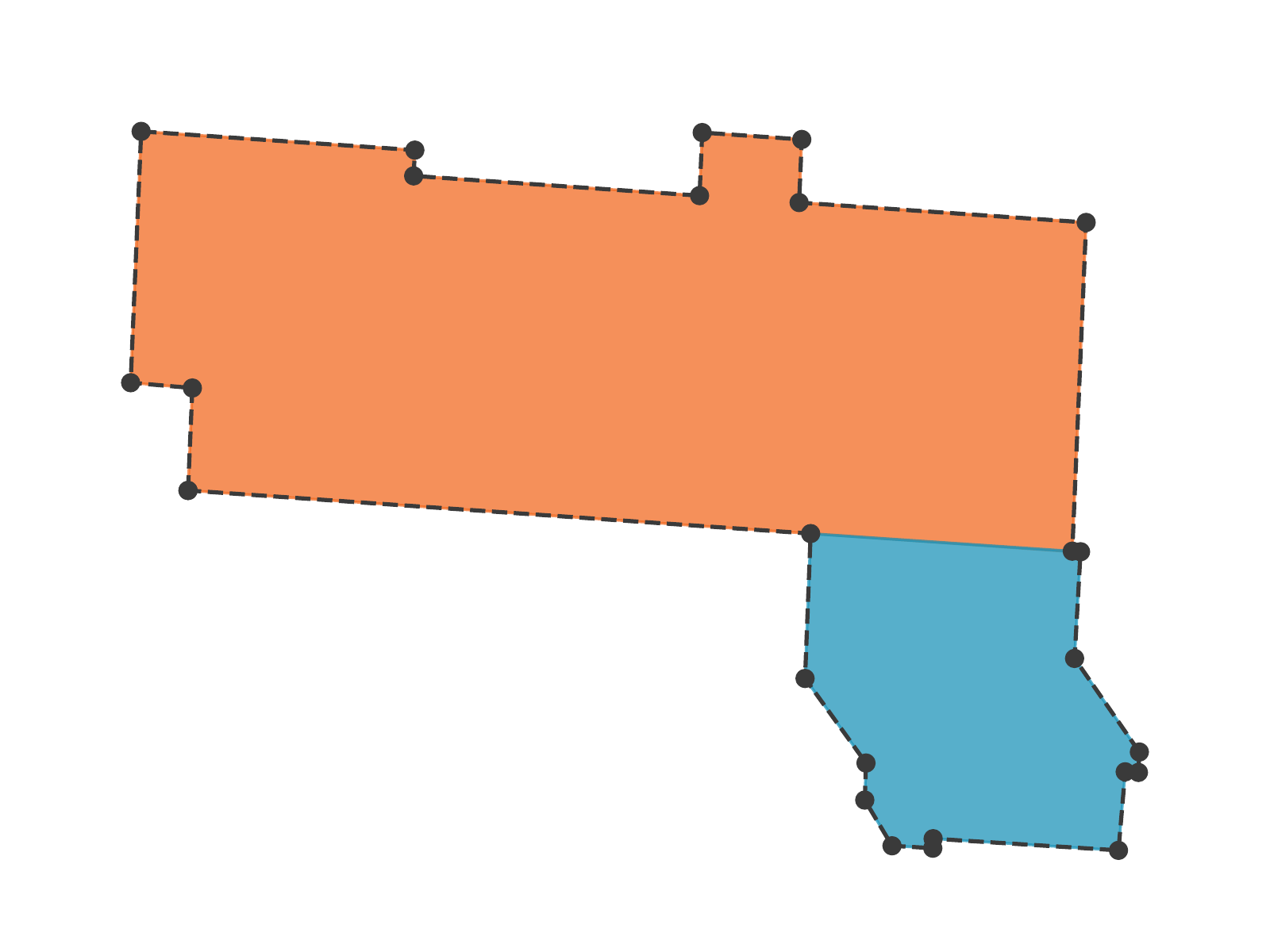}
    \label{fig:MergePolygons:simple:13}
  \end{subfigure}
  \caption{Examples of merged building footprints generated by
    Algorithm~\ref{alg:MergePolygons} for the Hammarkullen test
    case. In each panel, the two colored shapes show the two polygons
    to be merged and the dashed line and marked vertices show the
    result of the merging algorithm.}
  \label{fig:MergePolygons:real}
\end{figure}

A number of examples of merged polygons generated by
Algorithm~\ref{alg:MergePolygons} are provided in
Figure~\ref{fig:MergePolygons:simple} (simple test cases) and
Figure~\ref{fig:MergePolygons:real} (real-world data). The algorithm
handles the majority of building polygons very well but occasionally
fails in a few corner cases. It then falls back to computing the
convex hull of the two buildings, which is often a satisfactory
resolution.

\paragraph{Steps~2.2--2.4}

Step~2.1 potentially reduces the number buildings in the city model
when two or more buildings are merged into a single building. After
having completed Step~2.1, the city model is once again cleaned in
Step~2.2 (same as Step~1.4) and then all building heights are
recomputed; when two buildings are merged, their ground points and
roof points (computed in Step~1.6) are also merged and used for
(re)computing the heights in Step~2.3. Finally, the city model is
(re)exported to JSON format in Step~2.4.

\pagebreak

\subsection{Step 3: Generate city mesh}

After the completion of Step~2, the city model has been generated,
cleaned and simplified and is now suitable as input for the mesh
generation step. The key idea of the mesh generation is to first
generate a 2D mesh that respects the footprints of the building
footprints and then layer and transform the 2D mesh to create a 3D
volume mesh of the city model. The key steps in the mesh generation
process are outlined in Table~\ref{tab:step3}.

\begin{table}[htb]
  \centering
  \small
  \begin{tabular}{|l|l|}
    \hline
    Step~3.1 & Generate 2D mesh \\
    \hline
    Step~3.2 & Generate 3D mesh (layer 2D mesh) \\
    \hline
    Step~3.3 & Smooth 3D mesh (set ground height) \\
    \hline
    Step~3.4 & Trim 3D mesh (remove building interiors) \\
    \hline
    Step~3.5 & Smooth 3D mesh (set ground and building heights)  \\
    \hline
    Step~3.6 & Export mesh \\
    \hline
  \end{tabular}
  \caption{Essential substeps for Step~3: Generate city mesh.}
  \label{tab:step3}
\end{table}

\pagebreak

\paragraph{Step 3.1: Generate 2D mesh}

\begin{figure}[htb]
  \centering
  \begin{subfigure}[t]{0.475\textwidth}
    \centering
    \includegraphics[width=\figsize]{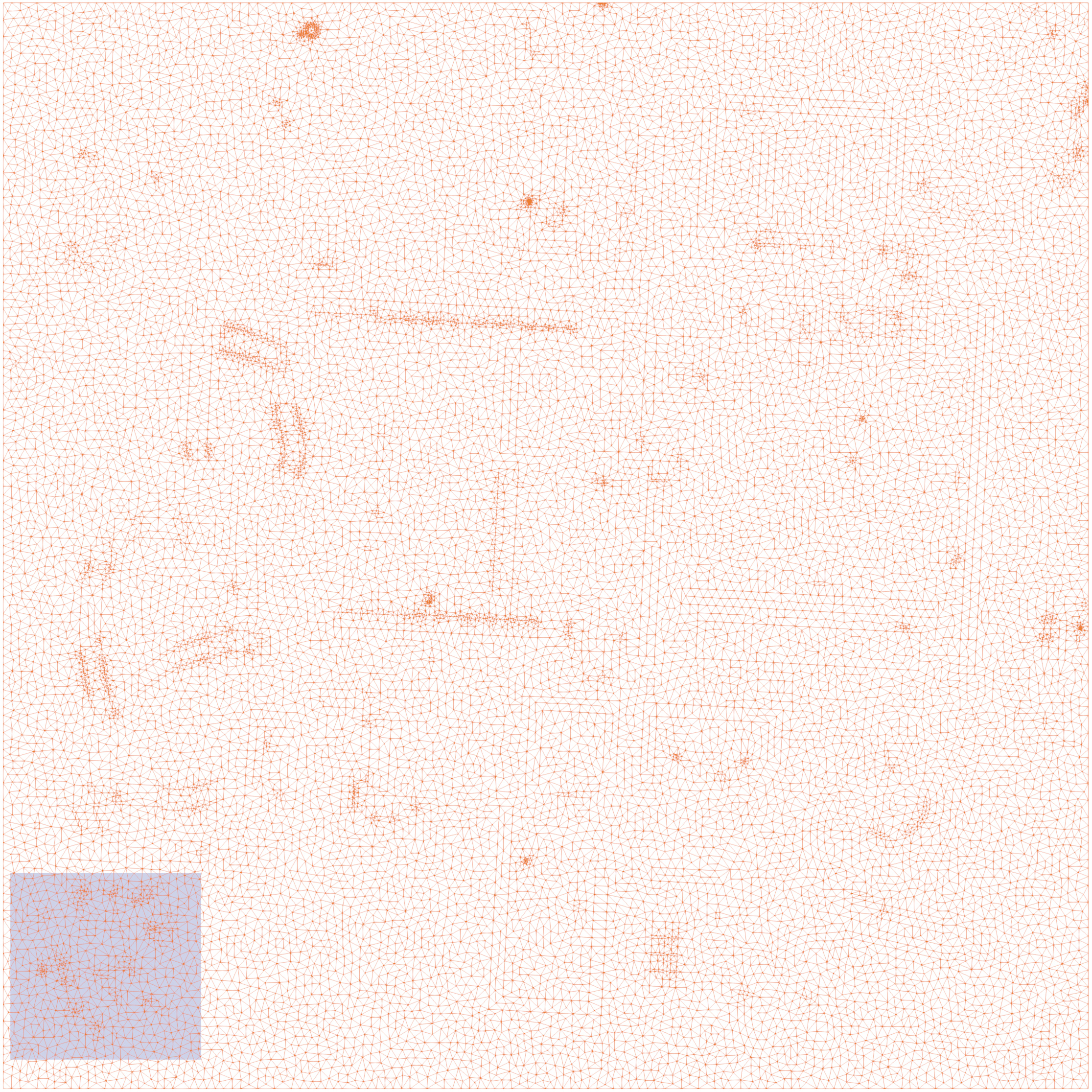}
    \caption{Full view}
    \label{fig:step3.1:full}
  \end{subfigure}
  \begin{subfigure}[t]{0.45\textwidth}
    \centering
    \includegraphics[width=\figsize]{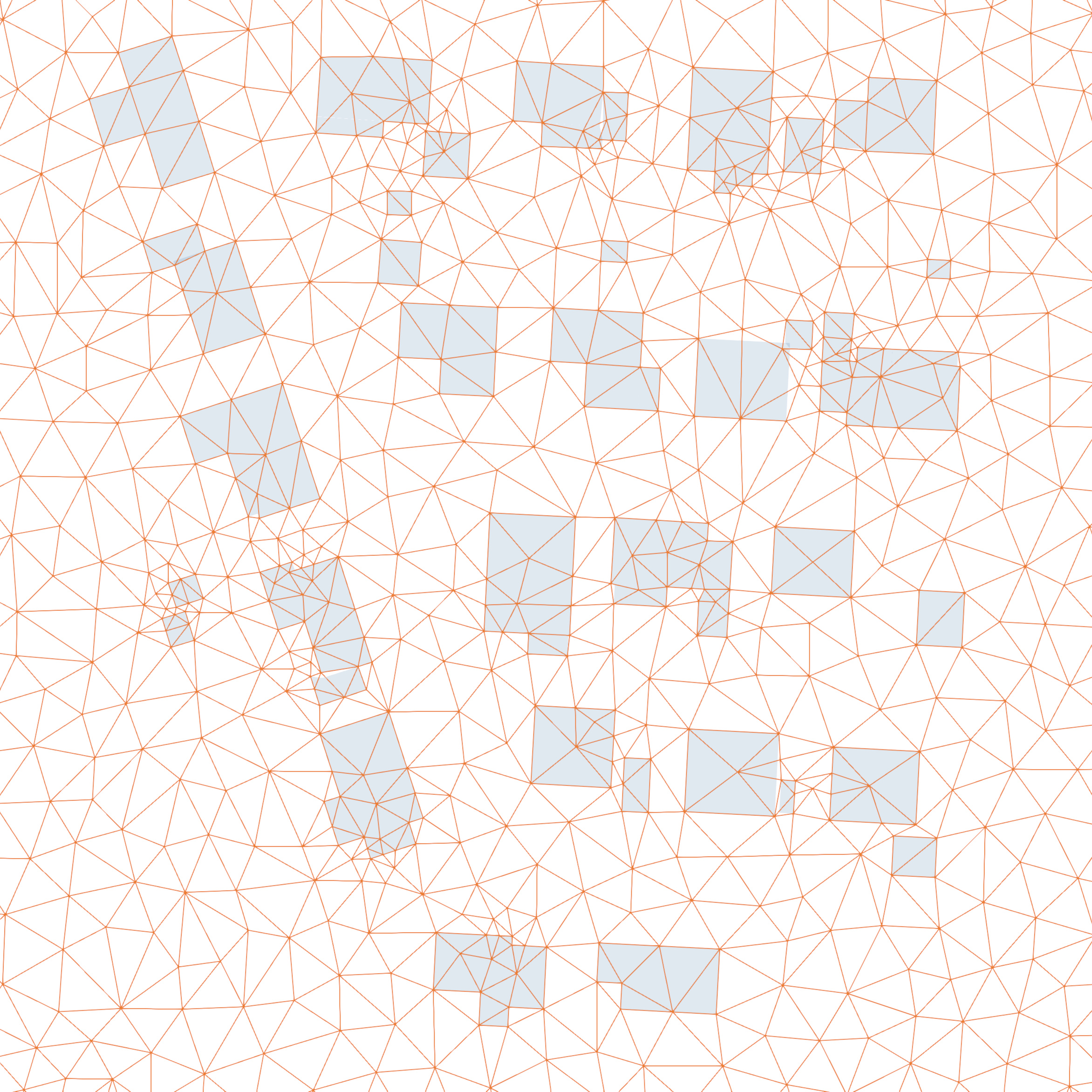}
    \caption{Detail}
    \label{fig:step3.1:detail}
  \end{subfigure}
  \caption{Output from Step~3.1 in the 3D mesh generation
    process for the Hammarkullen test case.}
  \label{fig:step3.1}
\end{figure}

A 2D mesh is generated of the computational domain (a bounding box
enclosing the city model) that respects the footprint of each building
of the (simplified) city model from Step~2. The mesh is generated by
including the edges of all building footprints as segments that must
be included in the triangular mesh generated by the 2D mesh
generator. The 2D mesh generator used is Triangle~\cite{shewchukTriangleEngineering2D1996},
which is a well-proven, robust and efficient 2D mesh generator that
generates triangular meshes of high quality. Figure~\ref{fig:step3.1}
illustrates the output of Step~3.1.

\paragraph{Step~3.2: Generate 3D mesh (layer 2D mesh)}

Once the 2D mesh is generated, a 3D mesh is created by layering the
2D mesh. The distance between the layers is set to the same mesh size
used for the 2D mesh generation in Step~3.1 and the number of layers
is determined by a given height of the computational domain.

The layering creates a 3D mesh of triangular prisms. Each such prism
is then divided into three tetrahedra. By following a consistent
numbering scheme of the triangles in the triangular mesh, we may
ensure that the generated 3D tetrahedral mesh is \emph{conforming};
that is, the intersection between each pair of tetrahedra is either
the empty set, a common vertex, a common edge, or a common face (no
hanging nodes).

\begin{figure}[htb]
  \centering
  \includegraphics[width=0.6\textwidth]{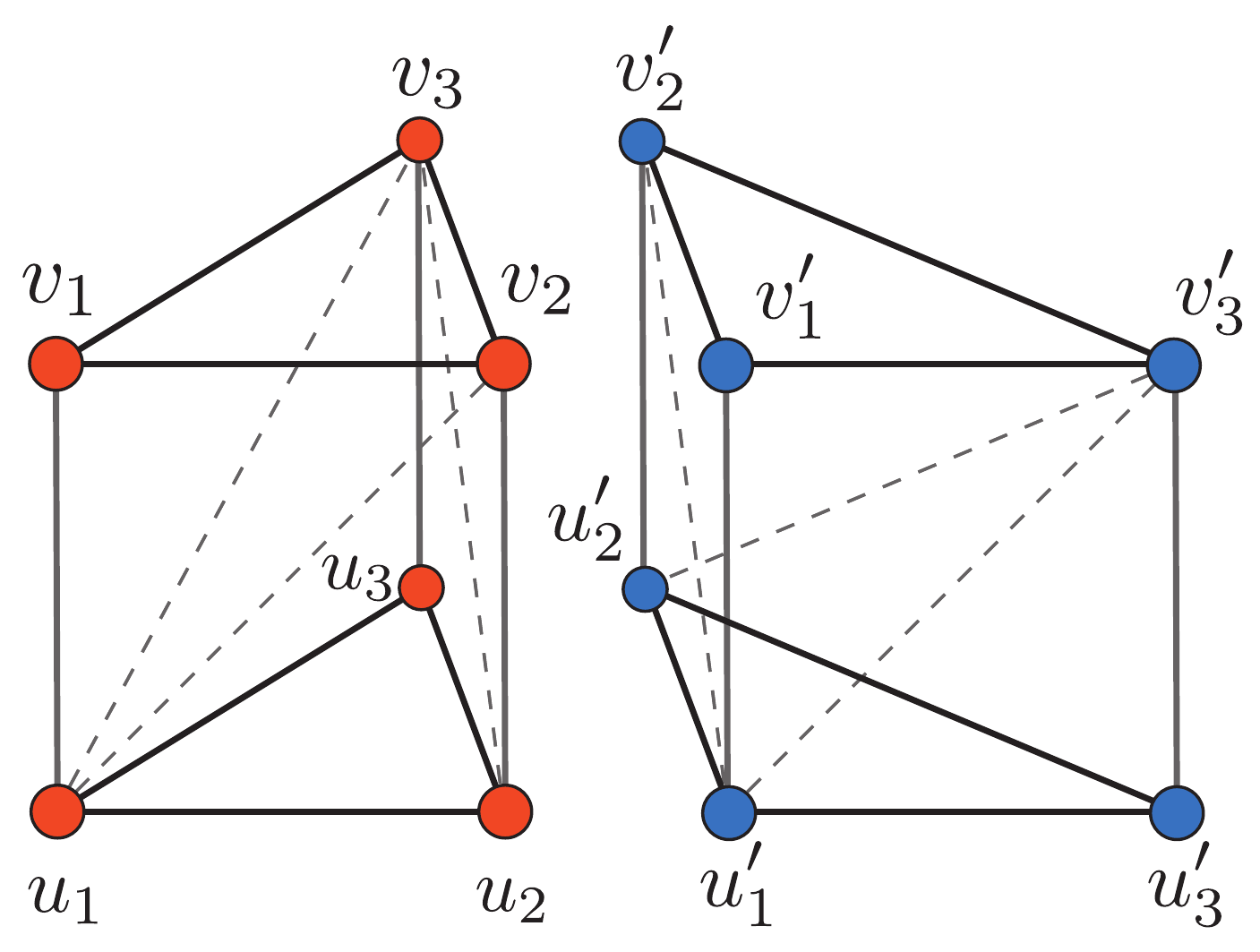}
  \caption{The triangular prisms created by layering the 2D mesh are
    divided into three tetrahedra in such a way that the tetrahedra
    generated from two neighboring prisms are always matching on the
    common boundary.}
  \label{fig:prismdivision}
\end{figure}

To be precise, let $u_1$, $u_2$, $u_3$ be the indices of the three
vertices that define a triangle in any given layer, and let $v_1$,
$v_2$, $v_3$ be the indices for the corresponding triangle in the next
layer. It is assumed that the vertices of each triangle are sorted by
global vertex index such that $u_1 < u_2 < u_3$ and $v_1 < v_2 <
v_3$. We then divide the prism defined by the two triangles into three
tetrahedra by
\begin{align}
  T_1 &= (u_1, u_2, u_3, v_3) \\
  T_2 &= (u_1, v_2, u_2, v_3) \\
  T_3 &= (u_1, v_1, v_2, v_3)
\end{align}
This numbering, in combination with the requirement that each triangle
be sorted by global vertex index, ensures that the tetrahedra created
from any two neighboring triangular prisms are matching on their
common boundary. See Figure~\ref{fig:prismdivision} for an
illustration.

\begin{figure}[htb]
  \centering
  \begin{subfigure}[t]{0.49\textwidth}
    \centering
    \includegraphics[width=\figsize]{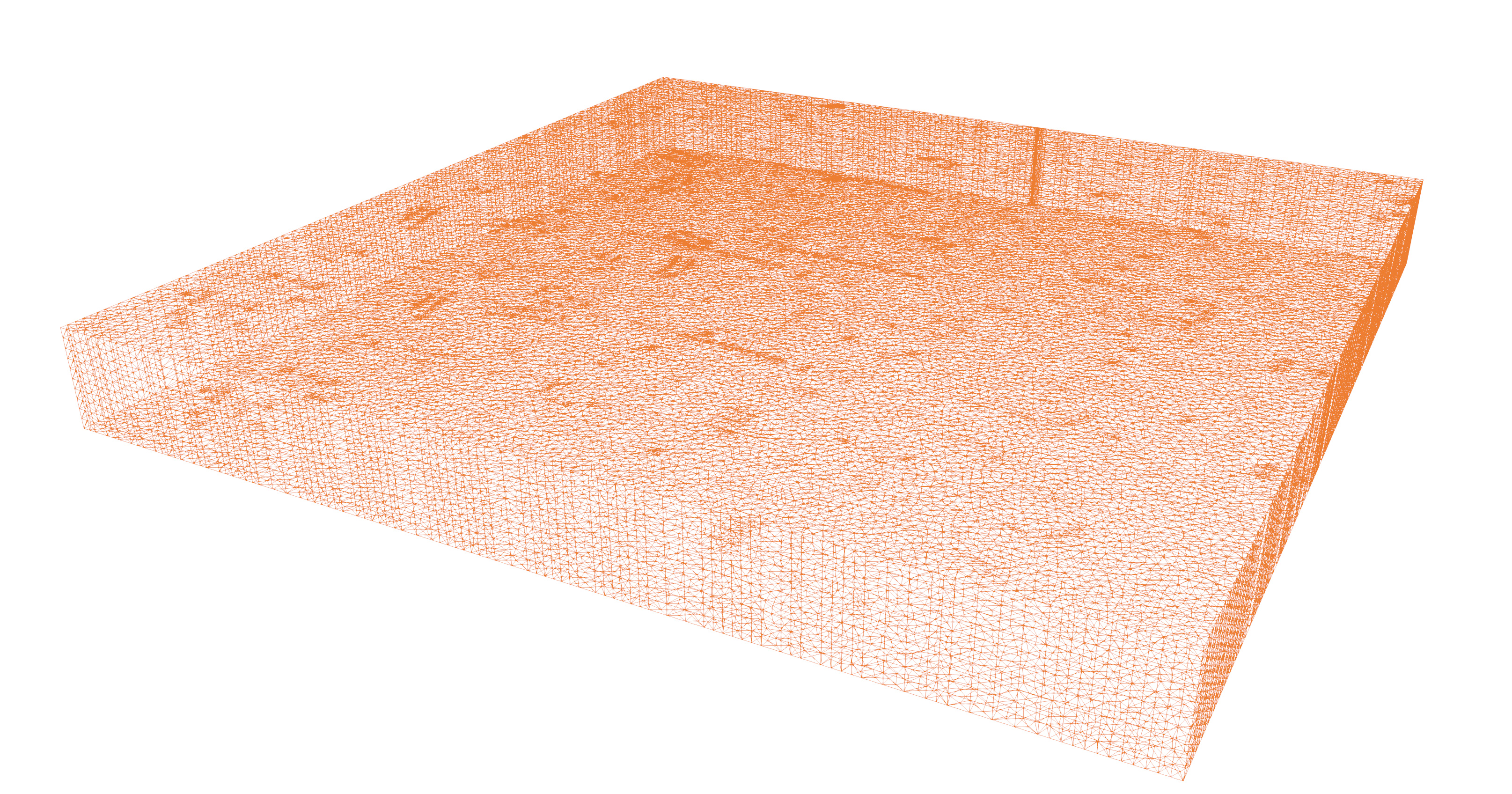}
    \caption{Step 3.2}
    \label{fig:step3.2}
  \end{subfigure}
  \begin{subfigure}[t]{0.49\textwidth}
    \centering
    \includegraphics[width=\figsize]{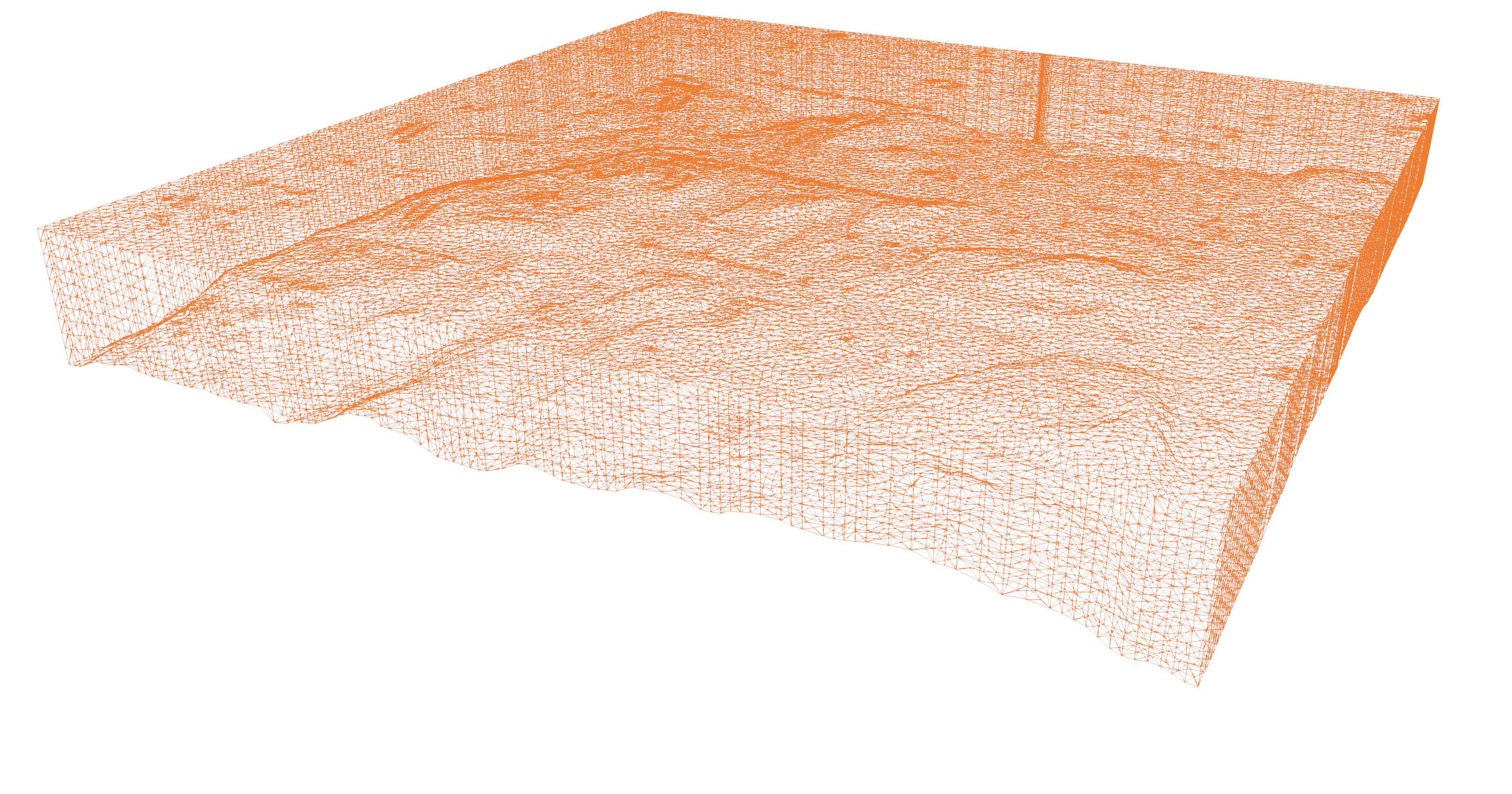}
    \caption{Step 3.3}
    \label{fig:step3.3}
  \end{subfigure}
  \begin{subfigure}[t]{0.49\textwidth}
    \centering
    \includegraphics[width=\figsize]{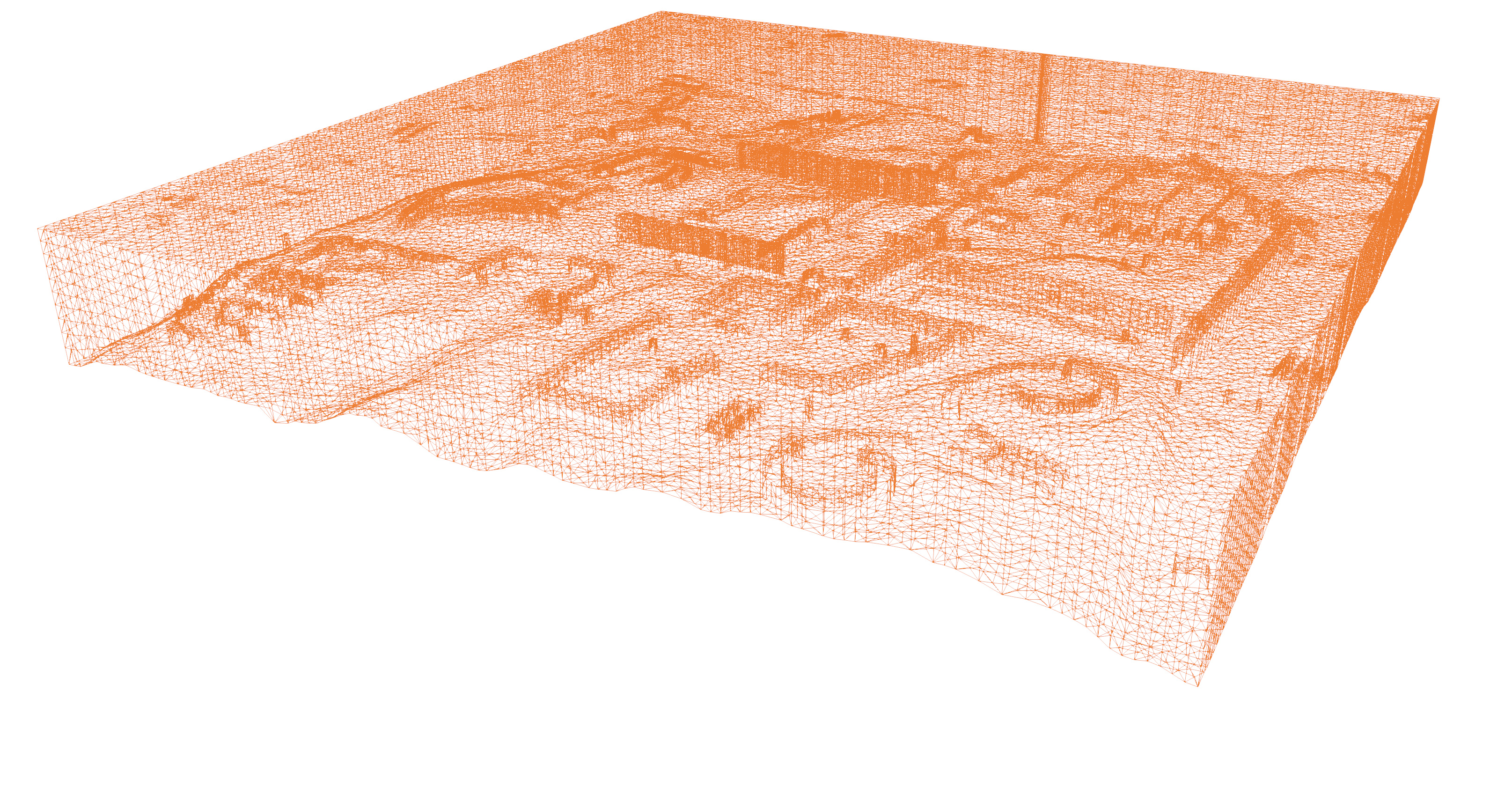}

    \caption{Step 3.4}
    \label{fig:step3.4}
  \end{subfigure}
  \begin{subfigure}[t]{0.49\textwidth}
    \centering
    \includegraphics[width=\figsize]{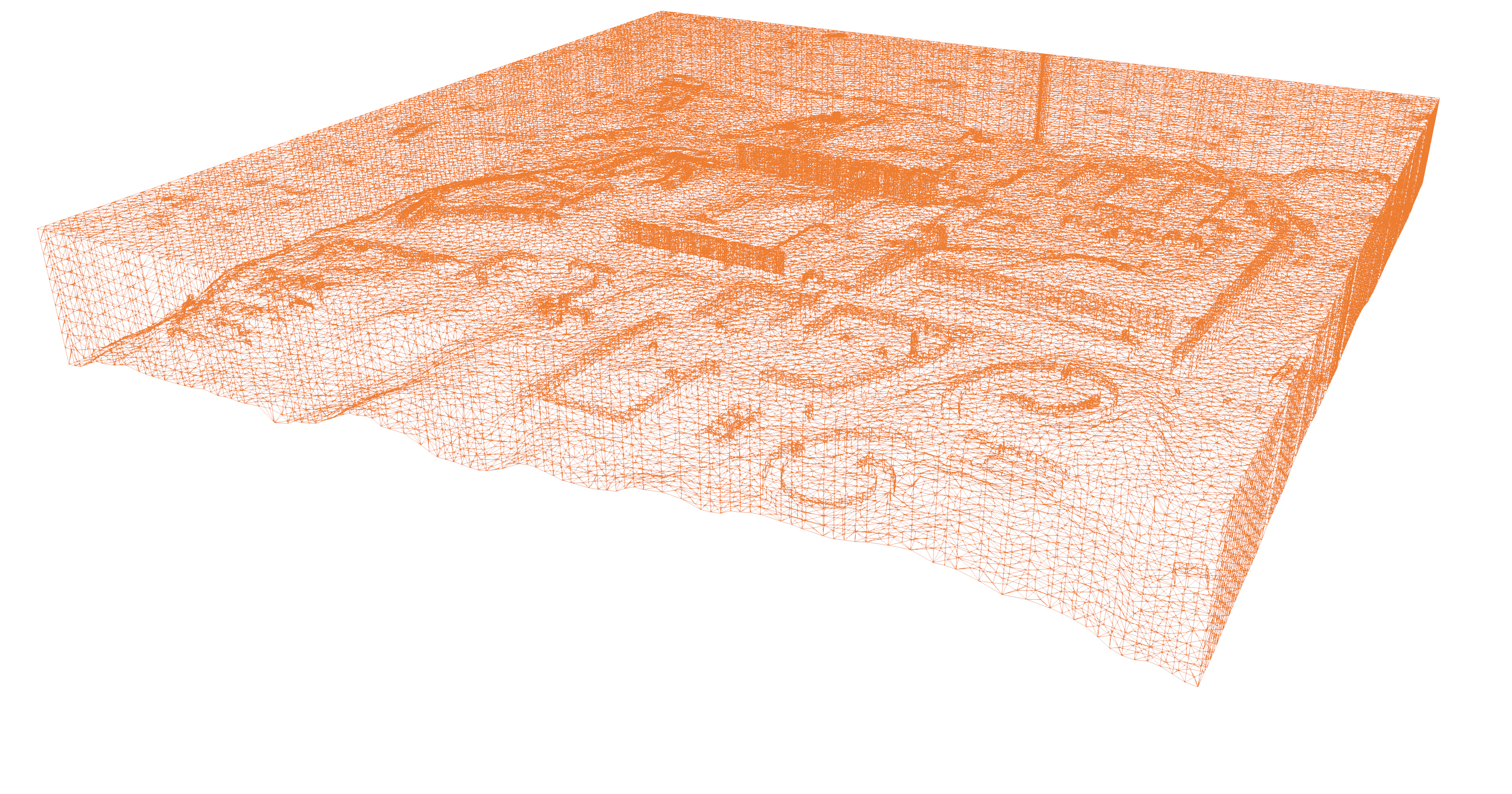}
    \caption{Step 3.5}
    \label{fig:step3.5}
  \end{subfigure}
  \caption{Output from Steps~3.2--3.5 in the 3D mesh generation
    process for the Hammarkullen test case.}
  \label{fig:steps3.2:3.5}
\end{figure}

The result of Step~3.2 is a tetrahedral (volume) 3D mesh of the
bounding box of the computational domain, where the tetrahedral faces
are ensured to respect the building boundaries. Figure~\ref{fig:step3.2}
illustrates the output of Step~3.2.

\paragraph{Step~3.3: Smooth 3D mesh (set ground height)}

The 3D mesh generated in Step~3.2 is next transformed to take the
topography of the city into account; that is, the mesh is transformed
to set the ground height as defined by the digital terrain map (DTM)
generated in Step~1.3. Note that we cannot simply displace the mesh
vertices at ground level to account for the ground height, since this
would (potentially and very likely) generate inverted tetrahedra by
pushing the vertices in the first layer past the vertices in the
second layer and beyond. Instead, we employ \emph{Laplacian mesh
  smoothing}~\cite{fieldLaplacianSmoothingDelaunay1988} with the ground height as a boundary
condition. This results in a smooth transformation of the mesh that
displaces the vertices in all layers, with the vertices in the first
layer set to the height given by the DTM.

\newcommand{\GG}{\Gamma_{\mathrm{G}}}
\newcommand{\GR}{\Gamma_{\mathrm{R}}}

Laplacian smoothing is based on solving a partial differential
equation (PDE), the solution of which defines the displacement of each
vertex of the mesh. To define the PDE, let $\Omega$ be the
computational domain defined by the 3D mesh, let $\GG$ be the ground
layer of the domain, and let $\mathrm{DTM} : \R^2\to\R$ be the
function defining the digital terrain map, defining the ground height
at any given point.

We solve the following boundary value problem for
the $z$-displacement of the 3D mesh:
\begin{align}
  -\Delta u &= 0 && \text{in }\Omega \\
  u(x, y, z) &= \mathrm{DTM}(x, y) - z && \text{on }\GG \\
  \partial_n u(x, y, z) &= 0 && \text{on }\partial\Omega\setminus\GG
\end{align}
Note that we thus solve a boundary value problem with the differential
ground height as a strong (Dirichlet) boundary condition at the bottom
of the computational domain, and homogeneous Neumann conditions on the
remaining boundary. The boundary value problem is defined and solved
using FEniCS~\cite{loggAutomatedSolutionDifferential2012} with linear ($P_1$) elements. The linear
system is solved using the BiCGSTAB Krylov method~\cite{vandervorstBiCGSTABFastSmoothly1992} and
an algebraic multigrid (AMG) preconditioner.

Once the solution has been computed, the $z$-displacement $u$ is added
to each vertex of the 3D mesh to obtain a transformed mesh that
respects the ground height. Figure~\ref{fig:step3.3} illustrates the
output of Step~3.3.

\pagebreak

\paragraph{Step~3.4: Trim 3D mesh (remove building interiors)}

The mesh obtained from Step~3.3 respects the ground height and the
faces of all tetrahedra respect the building boundaries. However, the
mesh still covers the entire computational domain, including the
interior of buildings. In Step~3.4, we trim the 3D mesh to remove all
tetrahedra from building interiors. Figure~\ref{fig:step3.4}
illustrates the output of Step~3.4.

\paragraph{Step~3.5: Smooth 3D mesh (set ground and building heights)}

The mesh obtained from Step~3.4 is almost complete. However, the
trimming process can only decide to either keep or remove
tetrahedra. As a result, the height of each building will be set to
the height of the closest layer, and the height of each building will
be correct only to within the mesh size used define the layer
distance. To obtain correct building heights, we once again perform
Laplacian smoothing as in Step~3.3, this time by solving the following
boundary value problem:
\begin{align}
  -\Delta u &= 0 && \text{in }\Omega \\
  u(x, y, z) &= \mathrm{DTM}(x, y) - z && \text{on }\GG \\
  u(x, y, z) &= \mathrm{BH}(x, y) - z && \text{on }\GR \\
  \partial_n u(x, y, z) &= 0 && \text{on }\partial\Omega\setminus(\GG\cup\GR)
\end{align}
where $\GR$ is the part of the boundary touching building roof tops
and where $\mathrm{BH} : \R^2\to\R$ is a function defining the
(absolute) building heights (computed in Step~2.3).

Note that in total two smoothing steps are performed, both before
(Step~3.2) and after (Step~3.4) trimming the mesh from tetrahedra
inside buildings. Two steps are essential to avoid the generation of
inverted tetrahedra; if the first smothing step is skipped and only
the second smoothing step is performed, then the second smoothing step
will be too agressive and will result in inverted tetrahedra in
regions close to building roof tops.

The result of Step~3.5 is the final tetrahedral 3D volume mesh of the
city model. The mesh is stored as a list of vertex positions and a
list of tetrahedral vertex indices as in~\cite{loggEfficientRepresentationComputational2009}. The mesh
discretizes the volume defined by subtracting from a bounding box of
the city model both the ground and the
buildings. Figure~\ref{fig:step3.5} illustrates the output of
Step~3.5.

\subsection{Data sources}

All data used in this study are public data obtained from the Swedish
Mapping, Cadastral and Land Registration Authority. In particular, we have used the dataset
"Lantmäteriet:\\Laserdata NH 2019" in LAZ format for point clouds and
the dataset "Lantmäteriet:\\Fastighetskartan bebyggelse 2020" in SHP
format for cadastral data (building footprints). The data are in
reference to the standard coordinate system for Sweden, EPSG:3006 also
known as SWEREF99 TM. The resolution of the point clouds used is
specified to between 0.5 and 1~points per square meter. Two different
datasets were used in this study, Majorna and Hammarkullen, two
regions in Gotheburg, Sweden. The dataset Majorna, used for the
benchmark study, is freely distributable from Lantmäteriet and can be
accessed through the Git repository of our
implementation~\cite{loggDTCCBuilder2022}. The dataset Hammarkullen, used for
illustrations in the Methods section, is public data but requires a
license from Lantmäteriet and may thus not be shared as part of our
implementation.

\pagebreak

\subsection{Implementation}

The algorithms described in this paper are implemented as part of the
open-source (MIT license) digital twin platform developed at the the
Digital Twin Cities Centre \\(\url{https://dtcc.chalmers.se}). In
particular, the software and data used to generate the results
presented in this paper are availalable as part of DTCC~Builder
version~\cite{loggDTCCBuilder2022}. The algorithms are implemented in
C++ and use several open-source packages, notably
FEniCS~\cite{loggAutomatedSolutionDifferential2012} for solving PDEs,
Triangle~\cite{shewchukTriangleEngineering2D1996} for 2D mesh
generation, and
GEOS~\cite{geoscontributorsGEOSCoordinateTransformation2021} for
geometric operations.

\section{Acknowledgments}

Both authors gratefully acknowledge valuable input from Dag Wästberg
on geodata processing, Jorge Gil and Nikos Pitsianis for insightful
discussions, and Orfeas Eleftheriou and Sanjay Somanath for valuable
assistance with this study. This work is part of the Digital Twin
Cities Centre supported by Sweden’s Innovation Agency Vinnova under
Grant No.  2019-00041.

\pagebreak

\bibliographystyle{unsrt}
\bibliography{bibliography}

\end{document}